\documentclass[aps,twocolumn,prc,showpacs]{revtex4}
\usepackage{mathrsfs}
\usepackage{longtable,lscape}
\usepackage{txfonts}
\usepackage{amssymb}
\usepackage{indentfirst}
\usepackage{graphicx,,booktabs}
\usepackage{color}
\usepackage{amssymb}
\usepackage{epsfig}

\newcommand{\vsig}{\mbox{\boldmath$\sigma$\unboldmath}}
\newcommand{\veps}{\mbox{\boldmath$\epsilon$\unboldmath}}
\newcommand{\valf}{\mbox{\boldmath$\alpha$\unboldmath}}

\newcommand{\be}{\begin{equation}}
\newcommand{\ee}{\end{equation}}
\newcommand{\bea}{\begin{eqnarray}}
\newcommand{\eea}{\end{eqnarray}}
\newcommand{\bean}{\begin{eqnarray*}}
\newcommand{\eean}{\end{eqnarray*}}

\newcommand{\gapproxeq}{\lower
.7ex\hbox{$\;\stackrel{\textstyle >}{\sim}\;$}}
\newcommand{\lapproxeq}{\lower
.7ex\hbox{$\;\stackrel{\textstyle <}{\sim}\;$}}

\begin{document}

\title{\textbf{ Radiative
transitions of charmonium states in a constituent quark model}}
\author{
Wei-Jun Deng, Li-Ye Xiao, Long-Cheng Gui, and Xian-Hui Zhong
\footnote {E-mail: zhongxh@hunnu.edu.cn} } \affiliation{ 1) Department
of Physics, Hunan Normal University, and Key Laboratory of
Low-Dimensional Quantum Structures and Quantum Control of Ministry
of Education, Changsha 410081, China }

\affiliation{ 2) Synergetic Innovation
Center for Quantum Effects and Applications (SICQEA),
Hunan Normal University,Changsha 410081,China}

\begin{abstract}
We study the electromagnetic (EM) transitions of the $nS$,
$nP$ ($n\leq 3$), and $nD$ ($n\leq 2$) charmonium states with a
constituent quark model. We obtain a reasonable description of
the EM transitions
of the well-established charmonium states $J/\psi$,
$\psi(2S)$, $\chi_{cJ}(1P)$, $h_c(1P)$ and $\psi(3770)$. We find that
the M2 transitions give notable corrections to
some E1 dominant processes by interfering with the E1 transitions. Our
predictions of EM decay properties for the higher
charmonium states are also presented and compared with other model
predictions. In particular, we discuss the EM decay properties
of some $``XYZ"$ states, such as $X(3823)$, $X(3872)$, $X(3915)$, $X(3940)$
and $X(4350)$ as conventional
charmonium states. Assuming $X(3872)$
as the $\chi_{c1}(2P)$ state, our predicted ratio $\Gamma[X(3872)\to
\psi(2S)\gamma]/\Gamma[X(3872)\to J/\psi \gamma]\simeq 4.0$ is
consistent with BaBar's measurement.
\end{abstract}


\maketitle

\section{Introduction}

During the past a few years, great progress has been made in the
observation of the
charmonia~\cite{Eichten:2007qx,Brambilla:2010cs,Bevan:2014iga}. From
the review of the Particle Data Group (PDG)~\cite{PDG}, one can see
that a fairly abundant charmonium spectroscopy has been established,
and many new charmonium-like $``XYZ"$ states above open-charm
thresholds have been discovered from experiments. The observations
of these new states not only deepen our understanding of the
charmonium physics, but also bring us many mysteries in this field
to be uncovered~\cite{Brambilla:2010cs,Voloshin:2007dx}. If these
newly observed $``XYZ"$ states are assigned as conventional
charmonium states, some aspects, such as measured mass and decay
properties are not consistent with the predictions. Thus, on one
hand we should test the validity of the previous theoretical models
in the descriptions of the new states, and at the same time develop
new approaches to study these new states. On the other hand, one
should consider these new charmonium-like $``XYZ"$ states as exotic
states and attempt to establish a new hadron
spectroscopy~\cite{Olsen:2014qna}.

Stimulated by the extensive progress made in the observation of the
charmonia, in this work we study the electromagnetic (EM)
transitions of charmonium in a constituent quark model. As we know,
the EM decays of a hadron are sensitive to its inner structure. The
study of the EM decays not only is crucial for us to determine the
quantum numbers of the newly observed charmonium states, but also
provides very useful references for our search for the missing
charmonium states in experiments. To deal with the EM decays, beside
the widely used potential
models~\cite{Godfrey:1985xj,Swanson:2005,Li:2009zu,Barnes:2003vb,Cao:2012du},
some other models, such as lattice
QCD~\cite{Dudek:2006ej,Dudek:2009kk,Chen:2011kpa,Becirevic:2012dc},
QCD sum rules~\cite{Khodjamirian:1979fa,Beilin:1984pf,Zhu:1998ih},
effective Lagrangian approach~\cite{DeFazio:2008xq,Wang:2011rt},
nonrelativistic effective field theories of
QCD~\cite{Brambilla:2005zw,Brambilla:2012be,Pineda:2013lta},
relativistic quark model~\cite{Ebert:2003}, relativistic Salpeter
method~\cite{Wang:2010ej}, light front quark model~\cite{Ke:2013zs},
and Coulomb gauge approach~\cite{Guo:2014zva} have been employed in
theory. Although some comparable predictions from different models
have been obtained, strong model dependencies still exist.

The constituent quark model used in present work has been well
developed and widely applied to meson photoproduction
reactions~\cite{Li:1995si,Li:1997gda, Zhao:2001kk, Saghai:2001yd,
Zhao:2002id, He:2008ty, He:2008uf, He:2010ii, Zhong:2011ti,
Zhong:2011ht, Xiao:2015gra,Zhao:1998fn,Zhao:1998rt}. Its recent
extension to describe the process of $\pi N$ and $KN$
scattering~\cite{Zhong:2007fx,
Zhong:2008km,Zhong:2013oqa,Xiao:2013hca} and investigate the strong
decays of baryons~\cite{Zhong:2007gp,Liu:2012sj,Xiao:2013xi} and
heavy-light mesons~\cite{Zhong:2008kd, Zhong:2009sk, Zhong:2010vq,
Xiao:2014ura} also turns out to be successful and inspiring. In this model,
to describe the EM transitions of a composite system
a special EM operator is adopted, in which the effects
of binding potential is included.
Furthermore, the possible higher EM multipole
contributions to a EM transition process can be included naturally.
We expect that our descriptions of the EM transitions of the heavy
quarkonium are more successful than those of strong decays of
heavy-light mesons or baryons. The reasons are that (i) there is no
light quark in the heavy quarkonium, thus, the relativistic effects
from the constituent quark are strongly suppressed; (ii) the
quark-photon EM coupling used in the EM transitions is well-defined
and model independent, however, the nonperturbative quark-meson
strong couplings used in the strong decay processes are still
effective interactions; (iii) the electric transitions are
independent on the heavy constituent quark mass, however, the strong
decay processes have some dependencies on the constituent quark
mass; (iv) it is natural to consider the emitted photon as a
pointlike particle in the EM transitions, however, it is only an
approximation to consider the emitted light pseudoscalar meson as a
pointlike particle in the strong decay processes.

The paper is organized as follows. In Sec.~\ref{spec}, a brief
review of charmonium spectroscopy in the constituent quark model is
given. In Sec.~\ref{EM}, an introduction of EM transitions described
in the constituent model is given. The numerical results are
presented and discussed in Sec.~\ref{RD}. Finally, a summary is
given in Sec.~\ref{sum}.

\section{charmonium spectroscopy} \label{spec}


For a quarkonium $Q\bar{Q}$ state, its flavor wave function should
have a determined $C$-parity. For a $C=+1$ state, its flavor wave
function is
\begin{eqnarray}
\Phi_{S}=\frac{1}{\sqrt{2}}(Q\bar{Q}+\bar{Q}Q),
\end{eqnarray}
while for a $C=-1$ state its flavor function is
\begin{eqnarray}
\Phi_{A}=\frac{1}{\sqrt{2}}(Q\bar{Q}-\bar{Q}Q).
\end{eqnarray}

The usual spin wave functions are adopted. For the spin $S=0$ state,
it is
\begin{eqnarray}
\chi^0_0&=&\frac{1}{\sqrt{2}}(\uparrow\downarrow-\downarrow\uparrow),
\end{eqnarray}
and for the spin $S=1$ states, the wave functions are
\begin{eqnarray}
\chi^1_1=\uparrow\uparrow, \ \ \chi^1_{-1}=\downarrow\downarrow, \ \
\chi^1_0=\frac{1}{\sqrt{2}}(\uparrow\downarrow+\downarrow\uparrow).
\end{eqnarray}

In this work, the space wave function of a charmonium state is
adopted by the nonrelativistic harmonic oscillator wave function,
i.e., $\psi^n_{L m}=R_{nL}Y_{L m}$. Where $n$ is the radial quantum
number, $L$ is the quantum number of relative orbital angular
momentum between quark and antiquark, and $m$ is the quantum number
of the third component of $L$. The detail of the space wave
functions can be found in our previous work~\cite{Zhong:2008kd}.


The total wave function of a $Q\bar{Q}$ system is a product of the
spin-, flavor-, spatial-, and color-wave functions, which should be
antisymmetric under the exchange of the two quarks for the
constraint of the generalized Pauli principle . The color wave function is always
symmetric, thus, the product of the spin-, flavor-, and spatial-wave
functions should be antisymmetric. The spectroscopy of some $S$-,
$P$-, and $D$-wave charmonium states classified in the constituent
quark model has been listed in Tabs.~\ref{tab1}.

\begin{table}[htb]
\begin{center}
\caption{ Charmonium states classified in the constituent quark
model. The Clebsch-Gordan series for the spin and angular-momentum
addition of the total wave function $|n ^{2S+1} L_{J} \rangle=
\sum_{m+S_z=J_z} \langle Lm,SS_z|JJ_z \rangle \psi^n_{Lm}
\chi_{S_z}\Phi $ has been omitted, where $\Phi_{A,S}$ is the flavor
wave function. $M_{\mathrm{exp}}$ stands for the experimental masses
(MeV) from observations, which are taken from the PDG~\cite{PDG}.
While $M_{\mathrm{NR}}$, $M_{\mathrm{GI}}$ and $M_{\mathrm{SNR}}$
stand for the predicted masses with NR, GI and SNR potential
models~\cite{Swanson:2005,Li:2009zu}. } \label{tab1}
\begin{tabular}{cccccccc}
\hline\hline
 $n^{2S+1}L_J$ & name &~~$J^{PC}$~~ &~~$M_{\mathrm{exp}}$~~  &~~$M_{\mathrm{NR/GI/SNR}}$~~ &Wave~function\\
 \hline
$1 ^3S_{1}$               &$J/\psi$        &$1^{--}$    &$3097$       &$3090/3098/3097$      &$\psi_{00}^0\chi_{S_{z}}^1 \Phi_{A}$\\
$1 ^1S_{0}$         &$\eta_{c}(1S)$        &$0^{-+}$    &$2984$       &$2982/2975/2979$      &$\psi_{00}^0\chi^0 \Phi_{S}$\\
$2 ^3S_{1}$             &$\psi(2S)$        &$1^{--}$    &$3686$       &$3672/3676/3673$      &$\psi_{00}^1\chi_{S_{z}}^1 \Phi_{A}$\\
$2 ^1S_{0}$         &$\eta_{c}(2S)$        &$0^{-+}$    &$3639$       &$3630/3623/3623$      &$\psi_{00}^1\chi^0 \Phi_{S}$\\
$3 ^3S_{1}$             &$\psi(3S)$        &$1^{--}$    &$4040$       &$4072/4100/4022$      &$\psi_{00}^2\chi_{S_{z}}^1 \Phi_{A}$\\
$3 ^1S_{0}$         &$\eta_{c}(3S)$        &$0^{-+}$    &$3940?$      &$4043/4064/3991$      &$\psi_{00}^2\chi^0 \Phi_{S}$\\
$1 ^3P_{2}$        &$\chi_{c2}(1P)$        &$2^{++}$    &$3556$       &$3556/3550/3554$      &$\psi_{1m}^0\chi_{S_{z}}^1 \Phi_{S}$\\
$1 ^3P_{1}$        &$\chi_{c1}(1P)$        &$1^{++}$    &$3511$       &$3505/3510/3510$      &$\psi_{1m}^0\chi_{S_{z}}^1 \Phi_{S}$\\
$1 ^3P_{0}$        &$\chi_{c0}(1P)$        &$0^{++}$    &$3415$       &$3424/3445/3433$      &$\psi_{1m}^0\chi_{S_{z}}^1 \Phi_{S}$\\
$1 ^1P_{1}$            &$h_{c}(1P)$        &$1^{+-}$    &$3525$       &$3516/3517/3519$      &$\psi_{1m}^0\chi^0 \Phi_{A}$\\
$2 ^3P_{2}$        &$\chi_{c2}(2P)$        &$2^{++}$    &$3927$       &$3972/3979/3937$      &$\psi_{1m}^1\chi_{S_{z}}^1 \Phi_{S}$\\
$2 ^3P_{1}$        &$\chi_{c1}(2P)$        &$1^{++}$    &$3872?$      &$3925/3953/3901$      &$\psi_{1m}^1\chi_{S_{z}}^1 \Phi_{S}$\\
$2 ^3P_{0}$        &$\chi_{c0}(2P)$        &$0^{++}$    &$3918?$      &$3852/3916/3842$      &$\psi_{1m}^1\chi_{S_{z}}^1 \Phi_{S}$\\
$2 ^1P_{1}$            &$h_{c}(2P)$        &$1^{+-}$    &             &$3934/3956/3908$      &$\psi_{1m}^1\chi^0 \Phi_{A}$\\
$3 ^3P_{2}$        &$\chi_{c2}(3P)$        &$2^{++}$    &$4350?$      &$4317/4337/4208$      &$\psi_{1m}^2\chi_{S_{z}}^1 \Phi_{S}$\\
$3 ^3P_{1}$        &$\chi_{c1}(3P)$        &$1^{++}$    &             &$4271/4317/4178$      &$\psi_{1m}^2\chi_{S_{z}}^1 \Phi_{S}$\\
$3 ^3P_{0}$        &$\chi_{c0}(3P)$        &$0^{++}$    &             &$4202/4292/4131$      &$\psi_{1m}^2\chi_{S_{z}}^1 \Phi_{S}$\\
$3 ^1P_{1}$            &$h_{c}(3P)$        &$1^{+-}$    &             &$4279/4318/4184$      &$\psi_{1m}^2\chi^0 \Phi_{A}$\\
$1 ^3D_{3}$         &$\psi_{3}(1D)$        &$3^{--}$    &             &$3806/3849/3799$      &$\psi_{2m}^0\chi_{S_{z}}^1 \Phi_{A}$\\
$1 ^3D_{2}$         &$\psi_{2}(1D)$        &$2^{--}$    &$3823$       &$3800/3838/3798$      &$\psi_{2m}^0\chi_{S_{z}}^1 \Phi_{A}$\\
$1 ^3D_{1}$             &$\psi_1(1D)$        &$1^{--}$  &$3778$       &$3785/3819/3787$      &$\psi_{2m}^0\chi_{S_{z}}^1 \Phi_{A}$\\
$1 ^1D_{2}$        &$\eta_{c2}(1D)$        &$2^{-+}$    &             &$3799/3837/3796$      &$\psi_{2m}^0\chi^0 \Phi_{S}$\\
$2 ^3D_{3}$         &$\psi_{3}(2D)$        &$3^{--}$    &             &$4167/4217/4103$      &$\psi_{2m}^1\chi_{S_{z}}^1 \Phi_{A}$\\
$2 ^3D_{2}$         &$\psi_{2}(2D)$        &$2^{--}$    &             &$4158/4208/4100$      &$\psi_{2m}^1\chi_{S_{z}}^1 \Phi_{A}$\\
$2 ^3D_{1}$             &$\psi_1(2D)$        &$1^{--}$  &$4191$       &$4142/4194/4089$      &$\psi_{2m}^1\chi_{S_{z}}^1 \Phi_{A}$\\
$2 ^1D_{2}$        &$\eta_{c2}(2D)$        &$2^{-+}$    &             &$4158/4208/4099$      &$\psi_{2m}^1\chi^0 \Phi_{S}$\\

\hline\hline

\end{tabular}
\end{center}
\end{table}

\section{EM transitions in the quark model}\label{EM}

In this section, we give an introduction of the model used in the
calculations. The quark-photon EM coupling at the tree level is
described by
\begin{eqnarray}
H_e=-\sum_j
e_{j}\bar{\psi}_j\gamma^{j}_{\mu}A^{\mu}(\mathbf{k},\mathbf{r})\psi_j,
\end{eqnarray}
where $\psi_j$ stands for the $j$-th quark field in a hadron. The
photon has three momentum $\mathbf{k}$, and the constituent quark
$\psi_j$ carries a charge $e_j$. Replacing the spinor $\psi_j$ by
$\psi_j^{\dag}$ so that the $\gamma$ matrices are replaced by the
matrix $\valf$, the EM transition matrix elements for a radiative
decay process can be written as~\cite{Li:1997gda}
\begin{eqnarray}
\mathcal{M}=\Big \langle f \Big | \sum_j e_{j}\valf_{j}\cdot\veps
e^{-i\mathbf{k}\cdot \mathbf{r}_{j}}   \Big | i  \Big \rangle,
\end{eqnarray}
where $|i \rangle$ and $| f \rangle $ stand for the initial and
final hadron states, respectively, and $\veps$ is the polarization
vector of the photon.

For a composite system, the relativistic Hamiltonian is taken to be
\begin{eqnarray}
\hat{H}=\sum_j (\valf_{j}\cdot \mathbf{p}_j +\beta_j
m_j)+\sum_{i,j}V(\mathbf{r}_i-\mathbf{r}_j).
\end{eqnarray}
By using the identity~\cite{Brodsky:1968ea,Li:1997gda}
\begin{eqnarray}
\valf_j\equiv i[\hat{H},\mathbf{r}_j],
\end{eqnarray}
we have
\begin{eqnarray}\label{eqa}
\mathcal{M}&=&i\Big\langle f \Big | [\hat{H},\sum_j e_{j}\mathbf{r}_{j}\cdot\epsilon e^{-i\mathbf{k}\cdot \mathbf{r}_{j}}]\Big | i \Big \rangle \nonumber \\
     &&+i\Big\langle f \Big | \sum_j e_{j}\mathbf{r}_{j}\cdot\veps\valf_{j}\cdot \mathbf{k} e^{-i\mathbf{k}\cdot \mathbf{r}_{j}}\Big| i \Big\rangle\nonumber \\
     &=&-i(E_{i}-E_{f}-\omega_{\gamma})\langle f | g_{e}| i \rangle-i\omega_{\gamma}\langle f | h_{e}| i\rangle,
\end{eqnarray}
with
\begin{equation}\label{he}
h_{e}=\sum_{j}e_{j}\mathbf{r}_{j}\cdot\veps (1-{\valf}_{j}\cdot
\hat{\mathbf{k}})e^{-i\mathbf{k}\cdot \mathbf{r}_{j}},
\end{equation}
and
\begin{equation}
g_{e}=\sum_{j}e_{j}\mathbf{r}_{j}\cdot\veps e^{-i\mathbf{k}\cdot
\mathbf{r}_{j}}.
\end{equation}
In Eq. (\ref{eqa}), $E_i$, $E_f$ and $\omega_\gamma$ stand for the
energies of the initial hadron state, final hadron state and emitted
photon, respectively. In the initial-hadron-rest system, we have
\begin{equation}
E_i=E_f+\omega_\gamma.
\end{equation}
Thus, we obtain
\begin{eqnarray}\label{amp0}
\mathcal{M}&=&-i\omega_{\gamma}\langle f | h_{e}| i \rangle.
\end{eqnarray}

In this model, the wave function of a hadron is adopted by the
nonrelativistic (NR) harmonic oscillator wave function, i.e.,
$\psi^n_{L m}=R_{nL}(r)Y_{L m}(\Omega)$. To match the NR wave functions of
hadrons, we should adopt the NR form of Eq.~(\ref{he}) in the
calculations. Following the procedures used in
~\cite{Brodsky:1968ea,Li:1997gda}, we obtain the NR expansion of
$h_e$ in Eq.(\ref{he}), which is given by
\begin{equation}\label{he2}
h_{e}\simeq\sum_{j}\left[e_{j}\mathbf{r}_{j}\cdot\veps-\frac{e_{j}}{2m_{j}
}\vsig_{j}\cdot(\veps\times\hat{\mathbf{k}})\right]e^{-i\mathbf{k}\cdot
\mathbf{r}_{j}}.
\end{equation}
It is interesting to find that the first and second terms in
Eq.(\ref{he2}) are responsible for the electric and
magnetic transitions, respectively. The second term
in Eq.(\ref{he2}) is the same as that used in
Ref.~\cite{Godfrey:1985xj}, while the first term in Eq.(\ref{he2})
differs from $(1/m_j) \mathbf{p}_j\cdot\veps$ used in Ref.~\cite{Godfrey:1985xj}
for the effects of the binding potential is included in the transition
operator.

Finally we can relate the standard helicity amplitude $\mathcal{A}$
of the EM decay process to the amplitude $\mathcal{M}$ in
Eq.(~\ref{amp0}) by the relation
\begin{eqnarray}\label{amp1}
\mathcal{A}&=&-i\sqrt{\frac{\omega_\gamma}{2}}\langle f | h_{e}| i
\rangle.
\end{eqnarray}
It is easily found that the helicity amplitudes for the electric and magnetic
transitions are
\begin{eqnarray}\label{amp2}
\mathcal{A}^{E}&=&-i\sqrt{\frac{\omega_\gamma}{2}}\Big \langle f
\Big | \sum_{j}e_{j}\mathbf{r}_{j}\cdot\veps e^{-i\mathbf{k}\cdot
\mathbf{r}_{j}} \Big | i \Big \rangle,\\
\mathcal{A}^{M}&=&+i\sqrt{\frac{\omega_\gamma}{2}}\Big \langle f
\Big | \sum_{j}\frac{e_{j}}{2m_{j}
}\vsig_{j}\cdot(\veps\times\hat{\mathbf{k}})e^{-i\mathbf{k}\cdot
\mathbf{r}_{j}}\Big | i\Big \rangle.
\end{eqnarray}


In the initial-hadron-rest system for the radiative decay precess,
the momentum of the initial hadron is $\bf{P}_i=0$, and that of the
final hadron state is $\bf{P}_f=-\textbf{k}$. Without losing generals,
we select the photon momentum along the $z$ axial ($\mathbf{k}=k\hat{\mathbf{z}}$), and
take the the polarization vector of the photon with the right-hand  form, i.e.,
$\veps=-\frac{1}{\sqrt{2}}(1,i,0) $, in our calculations.
To easily work out the EM transition matrix elements,
we use the multipole expansion of the plane wave
\begin{eqnarray}\label{amp3}
e^{-i\mathbf{k}\cdot\mathbf{r}_{j}}=e^{-ikz_j}=\sum_l \sqrt{4\pi (2l+1)}(-i)^lj_l(kr_j)Y_{l0}(\Omega),
\end{eqnarray}
where $j_l(x)$ is the Bessel function.
Then, we obtain the matrix element for the electric multipole transitions
with angular momentum $l$ (E$l$ transitions):
\begin{eqnarray}\label{amp4}
\mathcal{A}^{\mathrm{E}l}&=&\sqrt{\frac{\omega_\gamma}{2}}\Big \langle f
\Big | \sum_{j} (-i)^{l}B_l e_{j}r_{j} j_{l+1}(kr_j)Y_{l1}(\Omega) \Big | i \Big \rangle \nonumber\\
& &+\sqrt{\frac{\omega_\gamma}{2}}\Big \langle f
\Big | \sum_{j} (-i)^{l}B_l e_{j}r_{j} j_{l-1}(kr_j)Y_{l1}(\Omega) \Big | i \Big \rangle,
\end{eqnarray}
where $B_l\equiv\sqrt{\frac{2\pi l(l+1)}{2l+1}}$.
We also obtain the matrix element for the magnetic multipole transitions
with angular momentum $l$ (M$l$ transitions):
\begin{eqnarray}\label{amp5}
\mathcal{A}^{\mathrm{M}l}&=&\sqrt{\frac{\omega_\gamma}{2}}\Big \langle f
\Big | \sum_{j} (-i)^{l}C_l \frac{e_{j}\sigma_+}{2m_j} j_{l-1}(kr_j)Y_{l-1~0}(\Omega) \Big | i \Big \rangle,
\end{eqnarray}
where $C_l\equiv i\sqrt{8\pi (2l-1)}$, and $\sigma_+=\frac{1}{2}(\sigma_x+i\sigma_y)$ is the spin shift operator.
Obviously, the E$l$ transitions satisfy the parity selection rule: $\pi_i\pi_f=(-1)^l$;
while the M$l$ transitions satisfy the parity selection rule: $\pi_i\pi_f=(-1)^{l+1}$,
where $\pi_i$ and $\pi_f$ stand for the parities of the initial and final hadron states, respectively.

Finally, using the parity selection rules, one can express the EM helicity
amplitude $\mathcal{A}$ with the matrix elements of EM multipole transitions in a unified form:
\begin{eqnarray}\label{ampaa}
\mathcal{A}= \sum_{l}\Big\{ \frac{1+(-1)^{\pi_i\pi_f+l}}{2}\mathcal{A}^{\mathrm{E}l}
+ \frac{1-(-1)^{\pi_i\pi_f+l}}{2}\mathcal{A}^{\mathrm{M}l}\Big\},
\end{eqnarray}
which is consistent the standard multipole expansion in Ref.~\cite{Durand:1962zza}.
Combining the parity selection rules, we easily know the possible
EM multipole contributions to a EM transition considered in present work,
which are listed in Tab.~\ref{tabs}.

\begin{table}[htb]
\begin{center}
\caption{Possible
EM multipole contributions to a EM transition between two charmonium states. } \label{tabs}
\begin{tabular}{cccccccc}
\hline\hline
 process & multipole contribution  \\
 \hline
$n ^3S_{1}\longleftrightarrow m ^1S_{0}$               &M1  \\
$n ^3P_{J}\longleftrightarrow m ^3S_{1}$               &E1, M2  \\
$n ^1P_{1}\longleftrightarrow m ^1S_{0}$               &E1  \\
$n ^3D_{J}\longleftrightarrow m ^3P_{J}$               &E1, E3, M2, M4  \\
$n ^1D_{1}\longleftrightarrow m ^1P_{1}$               &E1, E3  \\
$n ^3P_{J}\longleftrightarrow m ^1P_{1}$               &M1, M3  \\
\hline\hline
\end{tabular}
\end{center}
\end{table}

With these helicity amplitudes $\mathcal{A}$ worked out according to
Eq.(\ref{ampaa}), one obtains the partial decay widths of the EM
transitions by
\begin{equation}\label{dww}
\Gamma=\frac{|\mathbf{k}|^2}{\pi}\frac{1}{2J_i+1}\frac{M_{f}}{M_{i}}\sum_{J_{fz},J_{iz}}|\mathcal{A}_{J_{fz},J_{iz}}|^2,
\end{equation}
where $J_i$ is the total  angular momenta of the initial mesons ,
$J_{fz}$ and $J_{iz}$ are the components of the total angular
momentum along the $z$ axis of initial and final mesons,
respectively. In the calculation, the standard parameters of the
quark model are adopted, no free parameters to be fitted. For the
oscillator parameter in the harmonic oscillator wave function we use
$\alpha=400$ MeV, and for the constituent $c$ quark mass we adopt
$m_{c}=1500$ MeV. For the fine-structure constant, we use
$\alpha_{\mathrm{EM}}\equiv e^2/(4\pi)=1/137$. For the
well-established charmonium states, their masses are adopted the
experimental average values from the PDG~\cite{PDG}. While for the
missing charmonium states, their masses are adopted from the
theoretical predictions.


\section{Results and discussions}\label{RD}

\subsection{$J/\psi\rightarrow \eta_c(1S) \gamma$}


The $J/\psi\rightarrow \eta_c(1S) \gamma$ is a typical M1 transition.
According to our calculation, we find that the
partial decay width of this process is proportional to
$\omega_\gamma^3/m_c^2$. We note that the constituent quark mass
$m_{c}\gg \omega_\gamma$, thus, the transition rate of
$J/\psi\rightarrow \eta_c(1S) \gamma$ is strongly suppressed by the
factor $\omega_\gamma^2/m_c^2$.

About the transition $J/\psi\rightarrow \eta_c(1S) \gamma$,
discrepancies still exist between the theoretical predictions and
experimental measurements. For example, the calculations from NR
potential model~\cite{Swanson:2005} and Coulomb gauge
approach~\cite{Guo:2014zva} give a large width
$\Gamma[J/\psi\rightarrow \eta_c(1S) \gamma]\simeq 2.9$ keV, which
is about a factor of 2 larger than the world average experimental
value $\Gamma(J/\psi\rightarrow \eta_c \gamma)\simeq 1.58\pm 0.37$
keV from the PDG~\cite{PDG}. In this work, our predicted partial
decay width for the $J/\psi\rightarrow \eta_c(1S) \gamma$ transition
is
\begin{eqnarray}
\Gamma[J/\psi\rightarrow \eta_c(1S) \gamma]\simeq 1.25 \
\mathrm{keV},
\end{eqnarray}
which is close to the world average data~\cite{PDG}, and also in
agreement with the values 1.05 keV and $(1.5\pm 1.0)$ keV predicted
in the frameworks of relativistic quark model~\cite{Ebert:2003} and
NR effective field theories of
QCD~\cite{Brambilla:2005zw,Pineda:2013lta}, respectively. However,
the recent measurements at KEDR give a large partial decay width
$\Gamma[J/\psi\rightarrow \eta_c(1S) \gamma]\simeq 2.98\pm
0.18^{+0.15}_{-0.33}$ keV~\cite{Anashin:2014wva}, which is
consistent with the lattice QCD results $2.4-2.9$
keV~\cite{Dudek:2006ej,Dudek:2009kk,Chen:2011kpa,Becirevic:2012dc},
the prediction of Coulomb gauge approach~\cite{Guo:2014zva}, and
that of NR and GI potential models~\cite{Swanson:2005} (see
Tab.~\ref{EM2}). Finally it should be mentioned that Li and Zhao
studied the intermediate hadronic meson loop contributions to the
$J/\psi\rightarrow \eta_c(1S) \gamma$ process, which might provide
explicit corrections to the M1 transition as
well~\cite{Li:2007xr,Li:2011ssa}. We hope more efforts on the
experimental side will be devoted to clarify the disagreement among
various experiments.

\subsection{Radiative transitions of $2S$ states}

\subsubsection{$\psi(2S)$}

The $\psi(2S)$ resonance, i.e., $\psi(3686)$, can decay into
$\chi_{cJ}(1P)\gamma$ and $\eta_c(1S,2S)\gamma$ channels, which have
been observed in experiments. The $\psi(2S)\to \chi_{cJ}(1P)\gamma$
decay processes are governed by the E1 transitions. While the
$\psi(2S)\to \eta_c(1S,2S)\gamma$ are typical M1 transitions.

Our predicted partial decay widths for the $\psi(2S)\to
\chi_{cJ}(1P)\gamma$ processes are
\begin{eqnarray}
\Gamma[\psi(2S)\to \chi_{c0}(1P)\gamma]& \simeq & 26 \ \mathrm{keV},\\
\Gamma[\psi(2S)\to \chi_{c1}(1P)\gamma]& \simeq & 22 \ \mathrm{keV},\\
\Gamma[\psi(2S)\to \chi_{c2}(1P)\gamma]& \simeq & 14 \ \mathrm{keV}.
\end{eqnarray}
From Tab.~\ref{EMPW1}, one can see that our results for the
$\psi(2S)\to \chi_{c0}(1P)\gamma, \chi_{c1}(1P)\gamma$ processes are
in good agreement with the experimental data, and the predictions of
the GI potential model~\cite{Swanson:2005} and relativistic quark
model~\cite{Ebert:2003}. However, our prediction of the partial
decay width $\Gamma[\psi(2S)\to \chi_{c2}(1P)\gamma]\simeq 14$ keV
is about a factor of 1.8 narrower than the world average
data~\cite{PDG}. It should be mentioned that according to the study
in Ref.~\cite{Zhao:2013jza}, the higher term corrections from the
coupled-channel effects of intermediate charmed mesons to the
$\psi(2S)\to \chi_{cJ}(1P)\gamma$ processes are negligibly small.

Furthermore, it is interesting to find that our predicted partial
width ratios
\begin{eqnarray}
\frac{\Gamma[\psi(2S)\to \chi_{c2}(1P)\gamma]}{\Gamma[\psi(2S)\to \chi_{c0}(1P)\gamma]}& \simeq &0.54,\\
\frac{\Gamma[\psi(2S)\to \chi_{c1}(1P)\gamma]}{\Gamma[\psi(2S)\to
\chi_{c0}(1P)\gamma]}& \simeq &0.85,
\end{eqnarray}
are in good agreement with the corresponding values $0.60$ and
$0.86$ predicted with NR potential models~\cite{Swanson:2005},
although their partial widths are larger than ours. We hope these
ratios can be measured in future experiments.

Finally, it should be mentioned that the M$2$ transitions can affect
the translations of $\psi(2S)\to \chi_{cJ}(1P)\gamma$ slightly
by interfering with the E$1$ transitions.
Including the effects of M$2$ transitions, we find that the
descriptions of the transitions of $\psi(2S)\to \chi_{c0}(1P)\gamma,
\chi_{c1}(1P)\gamma$ become better compared with the data (see
Tab.~\ref{EM2}).

For the typical M1 transitions $\psi(2S)\to \eta_c(1S,2S)\gamma$, our
predicted partial decay widths are
\begin{eqnarray}
\Gamma[\psi(2S)\to \eta_{c}(1S)\gamma]& \simeq & 2.41 \ \mathrm{keV},\\
\Gamma[\psi(2S)\to \eta_{c}(2S)\gamma]& \simeq & 0.10 \
\mathrm{keV}.
\end{eqnarray}
From Tab.~\ref{EM2}, we see that our results are close to the
average data from the PDG~\cite{PDG} and the predictions of
relativistic quark model~\cite{Ebert:2003}. The prediction of
$\Gamma[\psi(2S)\to \eta_{c}(1S)\gamma] \simeq 0.4(8)$ keV from
Lattice QCD is notably smaller than our result~\cite{Dudek:2009kk}.
However, the partial decay width of $\Gamma[\psi(2S)\to
\eta_{c}(1S)\gamma]$ is notably overestimated in the framework of GI
and NR potential models ~\cite{Swanson:2005}. It should be
emphasized that the intermediate hadronic meson loop might provide
explicit corrections to the $\psi(2S)\to \eta_c(1S,2S) \gamma$
process~\cite{Li:2007xr,Li:2011ssa}. More studies of the M1
transitions $\psi(2S)\to \eta_c(1S,2S)\gamma$ are needed in theory.


\subsubsection{$\eta_c(2S)$}

The $\eta_c(2S)$ resonance can decay into $h_c(1P)\gamma$ and
$J/\psi \gamma$ by the E1 and M1 transitions, respectively. Our
predicted partial decay width for the $\eta_c(2S)\to h_c(1P)\gamma$
process is
\begin{eqnarray}
\Gamma[\eta_c(2S)\to h_c(1P)\gamma]& \simeq & 18 \ \mathrm{keV}.
\end{eqnarray}
From Tab.~\ref{EMPW1}, it is seen that the predictions of the GI and
NR potential models~\cite{Swanson:2005} and relativistic quark
model~\cite{Ebert:2003} are about a factor of 2 larger than our
result. While our predicted partial decay width
\begin{eqnarray}
\Gamma[\eta_c(2S)\to J/\psi \gamma]& \simeq & 1.64 \ \mathrm{keV},
\end{eqnarray}
is in good agreement with the prediction of relativistic quark
model~\cite{Ebert:2003} (see Tab.~\ref{EM2}). The recent calculation
from Lattice QCD gives a fairly large width $\Gamma[\eta_c(2S)\to
J/\psi \gamma] =  (15.7\pm 5.7)$ keV~\cite{Becirevic:2014rda}, which
is about an order of magnitude larger than our prediction.
Furthermore, the calculations from potential models also give a
large width $\Gamma[\eta_c(2S)\to J/\psi \gamma] = (5.6\sim 7.9)$
keV~\cite{Swanson:2005}. The radiative transitions $\eta_c(2S)\to
h_c(1P)\gamma,J/\psi \gamma$ are still not measured in experiments.
For the very different predictions of these decays in theory, we
hope some observations can be carried out in future experiments.

\subsection{Radiative transitions of $1P$ states}

\subsubsection{$\chi_{cJ}(1P)$}

For the triplet $1P$ states $\chi_{cJ}(1P)$ ($J=0,1,2$), their main
radiative transitions are $\chi_{cJ}(1P)\to J/\psi \gamma$. In these
decay processes, except for the dominant E1 transitions, the M2
transitions are allowed as well. Neglecting
the effects of the M2 transitions, we find the partial decay widths
of $\Gamma[\chi_{cJ}(1P)\to J/\psi \gamma]$ are proportional to
$(\omega_\gamma^2/\alpha^2)\omega_\gamma$. We note that the photon
energy $\omega_\gamma$ is very close to the oscillator parameter
$\alpha$ (i.e., $\omega_\gamma\sim \alpha$), thus, one can easily
find that $\Gamma[\chi_{cJ}(1P)\to J/\psi \gamma]\sim
\omega_\gamma$. We calculate the partial decay widths
$\Gamma[\chi_{cJ}(1P)\to J/\psi \gamma]$, our results are
\begin{eqnarray}
\Gamma[\chi_{c0}(1P)\to J/\psi \gamma]& \simeq & 128 \ \mathrm{keV},\\
\Gamma[\chi_{c1}(1P)\to J/\psi \gamma]& \simeq & 275 \ \mathrm{keV},\\
\Gamma[\chi_{c2}(1P)\to J/\psi \gamma]& \simeq & 467 \ \mathrm{keV}.
\end{eqnarray}
From Tab.~\ref{EMPW1}, we find that our predictions of
$\Gamma[\chi_{c0,1}(1P)\to J/\psi \gamma]$ are in good agreement
with the world average data from the PDG~\cite{PDG}. While our
prediction of $\Gamma[\chi_{c2}(1P)\to J/\psi \gamma]$ is slightly
(a factor of 1.26 ) larger than the present average data $(371\pm
34)$ keV~\cite{PDG}. It is interesting to find that the E1
transitions of $\chi_{cJ}(1P)\to J/\psi \gamma$ predicted by us are
in good agreement with those of NR potential
model~\cite{Swanson:2005}. Furthermore, to understand the radiative
transition properties of $\chi_{c0,1}(1P)$ some studies were carried
out from Lattice QCD as well ~\cite{Dudek:2006ej,Chen:2011kpa},
however, good descriptions are still not obtained for some technical
problems.

It should be emphasized that the M2 transitions
have obvious corrections to the radiative transitions of
$\chi_{cJ}(1P)\to J/\psi \gamma$ by interfering with
the E1 transitions. In the $\chi_{c2}(1P)\to J/\psi \gamma$ process,
the M2 transition has a constructive interference with the E1 transition,
while in the $\chi_{c1,0}(1P)\to J/\psi \gamma$ processes,
the M2 transition has a destructive interference with the E1 transition.
The corrections from the M2 transition might reach to
$(10\sim 20)\%$ of the partial decay widths (see Tab.~\ref{EMPW1}).
Furthermore, the partial width ratios were also obviously affected by
the corrections from the M2 transition. For example, considering
the M2 contributions, we find that the partial width ratio of
\begin{eqnarray}
\frac{\Gamma[\chi_{c2}(1P)\to J/\psi \gamma]}{\Gamma[\chi_{c0}(1P)\to J/\psi \gamma]}\simeq 3.64
\end{eqnarray}
is notably larger than the value 2.54 without M2 contributions.
This ratio is expected to be tested by more precise measurements
in the future.

Recently, the coupled-channel effects of intermediate charmed mesons on the
radiative transitions $\chi_{cJ}(1P)\to J/\psi \gamma$ were also
studied with an effective Lagrangian approach in
Ref.~\cite{Zhao:2013jza}, the results show that the coupled-channel
effects on these decay processes are relatively weak.

\subsubsection{$h_{c}(1P)$}

For the singlet $h_{c}(1P)$ state, its main radiative transition is
$h_{c}(1P)\to \eta_c(1S) \gamma$. It is a typical E1
transition. The partial width of $\Gamma [h_{c}(1P)\to
\eta_c (1S)\gamma]$ is also proportional to
$(\omega_\gamma^2/\alpha^2)\omega_\gamma$. Thus, the width of $\Gamma
[h_{c}(1P)\to \eta_c(1S) \gamma]$ is the same order of magnitude as
$\Gamma[\chi_{cJ}(1P)\to J/\psi \gamma]$. Our predicted partial
decay width
\begin{eqnarray}
\Gamma[h_{c}(1P)\longrightarrow \eta_{c} \gamma]\simeq 587 \
\mathrm{keV},
\end{eqnarray}
is in good agreement with the predictions from the relativistic
quark model~\cite{Ebert:2003} and Lattice QCD~\cite{Dudek:2006ej},
and also consistent with the data within its large
uncertainties~\cite{PDG} (see table~\ref{EMPW1}). The NR and GI
potential models and Coulomb gauge approach~\cite{Guo:2014zva} give
a notably narrower width than ours~\cite{Swanson:2005}. It should be
mentioned that the recent lattice calculation gave a larger width of
$\Gamma[h_{c}(1P)\to \eta_{c} \gamma]=720(70)$ keV. Similar result
was also obtained with a light front quark model~\cite{Ke:2013zs}.
As a whole, there are large discrepancies between different model
predictions of $\Gamma [h_{c}(1P)\to \eta_c \gamma]$. However, the
present world data can not be used to test the various predictions
for their large uncertainties. More accurate measurements for the
transition $h_{c}(1P)\to \eta_c \gamma$ are needed.

Finally, we give an estimation of partial decay width for
the M1 transition $h_{c}(1P)\to \chi_{c0}(1P) \gamma$. Our result is
\begin{eqnarray}
\Gamma[h_{c}(1P)\longrightarrow \chi_{c0}(1P) \gamma]=0.39 \
\mathrm{keV}.
\end{eqnarray}
The corresponding branching ratio is
\begin{eqnarray}
\mathcal{B}[h_{c}(1P)\longrightarrow \chi_{c0}(1P) \gamma]\simeq 5.6
\times 10^{-4}.
\end{eqnarray}
This sizeable transition rate indicates that the M1 transition
$h_{c}(1P)\to \chi_{c0}(1P) \gamma$ might be observed in forthcoming
experiments.

\subsection{Radiative transitions of $1D$ states}

\subsubsection{$\psi(3770)$}

The $\psi(3770)$ resonance is primarily a $\psi_1(1D)$ state with
small admixtures of $\psi(2S)$~\cite{Eichten:2007qx}. It can decay
into $\chi_{cJ}(1P)\gamma$. These decay processes are dominated by the E1
transition. Furthermore, the higher order E3, M2 and M4 transitions are allowed as well.

Considering $\psi(3770)$ as a pure
$\psi_1(1D)$ state, we calculate the radiative decay widths of
$\Gamma[\psi(3770)\to \chi_{cJ}(1P)\gamma]$. Our results
\begin{eqnarray}
\Gamma[\psi(3770)\to \chi_{c0}(1P)\gamma]& \simeq & 218 \ \mathrm{keV},\\
\Gamma[\psi(3770)\to \chi_{c1}(1P)\gamma]& \simeq & 70 \ \mathrm{keV},\\
\Gamma[\psi(3770)\to \chi_{c2}(1P)\gamma]& \simeq & 2.6 \
\mathrm{keV},
\end{eqnarray}
are consistent with the world average data from PDG~\cite{PDG}, and
also very close to the predictions of GI model~\cite{Swanson:2005}
(see Tab.~\ref{D-wave}). It is interesting to find that recently
the BESIII Collaboration gave their precise measurement of the partial
width of $\Gamma[\psi(3770)\to \chi_{c1}(1P)\gamma]=67.5\pm 10.8$
keV~\cite{Ablikim:2015sol}, which is in good agreement with our
prediction. Furthermore, our predicted partial width ratio
\begin{eqnarray}
\frac{\Gamma[\psi(3770)\to \chi_{c0}(1P)\gamma]}{\Gamma[\psi(3770)\to \chi_{c1}(1P)\gamma)]}& \simeq &3.1, 
\end{eqnarray}
is in good agreement with the experimental data $2.5\pm 0.6$ from
CLEO Collaboration as well~\cite{Briere:2006ff}.

It should be emphasized that the M2
transitions also have an obvious contribution [about $(10\sim
15)\%$] to the radiative transitions of $\Gamma[\psi(3770)\to
\chi_{cJ}(1P) \gamma]$ by interfering with the E$1$ transitions.

Finally, it should be mentioned that the
present measurements for the radiative transitions of $\psi(3770)\to
\chi_{cJ}(1P)\gamma$ are not precise enough to determine the small
mixing angle between $\psi_1(1D)$ and $\psi(2S)$.

\subsubsection{$X(3823)$}

Recently, $X(3823)$ as a good candidate for $\psi_2(1D)$ was
observed by the Belle Collaboration in the $B\to \chi_{c1}\gamma K$
decay with a statistical significance of
$3.8\sigma$~\cite{Bhardwaj:2013rmw}. Lately, this state was
confirmed by the BESIII Collaboration in the process $e^+e^-\to
\pi^+\pi^-X(3823)\to \pi^+\pi^-\chi_{c1}\gamma$ with a statistical
significance of $6.2\sigma$~\cite{Ablikim:2015dlj}.

Considering $X(3823)$ as the $\psi_2(1D)$ state, we predict the radiative decay
widths of $\Gamma[X(3823)\to \chi_{cJ}(1P)\gamma]$. Our results are
\begin{eqnarray}
\Gamma[X(3823)\to \chi_{c0}(1P)\gamma]& \simeq & 1.42 \ \mathrm{keV},\\
\Gamma[X(3823)\to \chi_{c1}(1P)\gamma]& \simeq & 227 \ \mathrm{keV},\\
\Gamma[X(3823)\to \chi_{c2}(1P)\gamma]& \simeq & 42 \ \mathrm{keV}.
\end{eqnarray}
Where we can find that the radiative transition rate of $X(3823)\to
\chi_{c0}(1P)\gamma$ is relatively small. The radiative transitions
of $X(3823)$ are dominated by the $\chi_{c1}(1P)\gamma$ channel.
This can explain why the $X(3823)$ is firstly observed in the
$\chi_{c1}(1P)\gamma$ channel, while not observed in the
$\chi_{c0,c2}(1P)\gamma$ channels. It should be mentioned that the
M2 transitions could give a $\sim15\%$ correction to
$\Gamma[X(3823)\to \chi_{c2}(1P)\gamma]$ by interfering
with the E1 transitions. Furthermore, Our predicted partial width ratio,
\begin{eqnarray}
\frac{\Gamma[X(3823)\to \chi_{c2}(1P)\gamma]}{\Gamma[X(3823)\to
\chi_{c1}(1P)\gamma]}& \simeq &19\%,
\end{eqnarray}
is consistent with the observations $< 42\%$~\cite{Ablikim:2015dlj}.
Our predictions of the $\Gamma[X(3823)\to \chi_{c1,c2}(1P)\gamma]$
are compatible with the other theoretical
predictions~\cite{Qiao:1996ve,Ebert:2003,Swanson:2005,Eichten:2002qv}.

According to the calculations from various models, the partial width
of $\Gamma[X(3823)\to \chi_{c2}(1P)\gamma]$ is not small. Thus,
looking for $X(3823)$ in the $\chi_{c2}(1P)\gamma$ channel and an
accurate measurement of the ratio $\Gamma[X(3823)\to
\chi_{c2}(1P)\gamma]/\Gamma[X(3823)\to \chi_{c1}(1P)\gamma]$ are
crucial to further confirm $X(3823)$ as the $\psi_2(1D)$ state.

\subsubsection{$\psi_3(1D)$ and $\eta_{c2}(1D)$}

Another two $1D$-wave states $\psi_3(1D)$ and $\eta_{c2}(1D)$ have
not been observed in experiments. According to theoretical
predictions, their masses are very similar to that of $\psi_2(1D)$.
If $X(3823)$ corresponds to the $\psi_2(1D)$ state indeed, the
masses of the $\psi_3(1D)$ and $\eta_{c2}(1D)$ resonances should be
around $3.82$ GeV.

For the singlet $1D$ state $\eta_{c2}(1D)$, its main radiative
transition is $\eta_{c2}(1D)\to h_c(1P) \gamma$. This process
is governed by the E1 transition, the effects from the E3
transition are negligibly small. Taking the mass of
$\eta_{c2}(1D)$ with $M=3820$ MeV, we predict the partial decay
width
\begin{eqnarray}
\Gamma[\eta_{c2}(1D)\to h_c(1P) \gamma]& \simeq & 261 \
\mathrm{keV}.
\end{eqnarray}
Our result is compatible with that from the relativistic quark
model~\cite{Ebert:2003} and effective Lagrangian
approach~\cite{DeFazio:2008xq}. The predictions from the potential
models~\cite{Swanson:2005,Li:2009zu} are slightly larger than ours
(see Tab.~\ref{D-wave}).

While for the triplet $1D$ state $\psi_3(1D)$, its common radiative
transitions are $\psi_3(1D)\to \chi_{cJ}(1P)\gamma$. In these
transitions, E1, E3, M2, M4 decays are allowed. Taking the mass of
$\psi_3(1D)$ with $M=3830$ MeV, we calculate the partial decay
widths $\Gamma[\psi_{3}(1D)\to \chi_{cJ}(1P) \gamma]$. Our results
are
\begin{eqnarray}
\Gamma[\psi_{3}(1D)\to \chi_{c0}(1P)\gamma]& \simeq & 0.87 \ \mathrm{keV},\\
\Gamma[\psi_{3}(1D)\to \chi_{c1}(1P)\gamma]& \simeq & 0.82 \ \mathrm{keV},\\
\Gamma[\psi_{3}(1D)\to \chi_{c2}(1P)\gamma]& \simeq & 226 \
\mathrm{keV}.
\end{eqnarray}
The interferences between M2 and E3 transitions are responsible for
the tiney partial decay widths of $\Gamma[\psi_{3}(1D)\to
\chi_{c0,c1}(1P) \gamma]$, while the E1 transitions govern the partial width of
$\Gamma[\psi_{3}(1D)\to \chi_{c2}(1P) \gamma]$. The magnitude of the
partial decay width of $\Gamma[\psi_{3}(1D)\to \chi_{c2}(1P)
\gamma]$ predicted by us is compatible with that from the potential
models~\cite{Swanson:2005,Li:2009zu} and the relativistic quark
model~\cite{Ebert:2003}. We hope the experimental Collaboration can
carry out a search for the missing $\psi_{3}(1D)$ state in the
$\chi_{c2}(1P) \gamma$ channel.

\subsection{Radiative transitions of $2P$ states} 

\subsubsection{$\chi_{c2}(2P)$}

In the $2P$ charmonium states, only the $\chi_{c2}(2P)$ has been
established experimentally. This state was observed by both
Belle~\cite{Uehara:2005qd} and BaBar~\cite{Aubert:2010ab} in the
two-photon fusion process $\gamma\gamma\to D\bar{D}$ with a mass
$M\simeq 3927$ MeV and a narrow width $\Gamma\simeq 24$
MeV~\cite{PDG}. We analyze its radiative transitions to $\psi(1
D)\gamma$, $J/\psi\gamma$ and $\psi(2S)\gamma$. We find the
radiative transition rate to $\psi(1 D)\gamma$ is very weak. The
predicted partial decay widths are
\begin{eqnarray}
\Gamma[\chi_{c2}(2P)\to \psi(3770)\gamma]& \simeq & 0.32 \
\mathrm{keV},\\
\Gamma[\chi_{c2}(2P)\to \psi_2(1D)\gamma]& \simeq & 1.5 \
\mathrm{keV},\\
\Gamma[\chi_{c2}(2P)\to \psi_3(1D)\gamma]& \simeq & 6.3 \
\mathrm{keV}.
\end{eqnarray}
In the calculation we take the mass of 3830 MeV for $\psi_3(1D)$.
Our predictions of the $\Gamma[\chi_{c2}(2P)\to \psi(3770)\gamma]$
and $\Gamma[\chi_{c2}(2P)\to \psi_2(1D)\gamma]$ are roughly
compatible with the results from the GI potential
model~\cite{Swanson:2005} (see Tab.~\ref{EMPW1}). However, our
prediction of the $\Gamma[\chi_{c2}(2P)\to \psi_3(1D)\gamma]$ is
about an order of magnitude smaller than that from the GI and NR
potential models~\cite{Swanson:2005} (see Tab.~\ref{EMPA}). Our
predictions indicate that it might be a challenge to look for the
missing charmonium state $\psi_3(1D)$ via the radiative transition
$\chi_{c2}(2P)\to \psi_3(1D)\gamma$.

The $\chi_{c2}(2P)$ state might have large radiative decay rates
into $J/\psi\gamma$ and $\psi(2S)\gamma$. In our calculations, we
find
\begin{eqnarray}
\Gamma[\chi_{c2}(2P)\to J/\psi\gamma]& \simeq & 34 \
\mathrm{keV},\\
\Gamma[\chi_{c2}(2P)\to \psi(2S)\gamma]& \simeq & 133 \
\mathrm{keV}.
\end{eqnarray}
It should be mentioned that the M2 transitions could give a $\sim
10-15\%$ correction to these radiative decay widths by interfering
with the E1 transitions. The branching fractions of these two radiative decay process could
reach to
\begin{eqnarray}
\mathcal{B}[\chi_{c2}(2P)\to J/\psi\gamma]& \simeq & 1.4\times
10^{-3} \
\mathrm{keV},\\
\mathcal{B}[\chi_{c2}(2P)\to \psi(2S)\gamma]& \simeq & 5.5\times
10^{-3} \ \mathrm{keV}.
\end{eqnarray}
Thus, the radiative transitions $\chi_{c2}(2P)\to
\psi(2S)\gamma,J/\psi\gamma$ are most likely to be observed in
forthcoming experiments.

Finally, it should be pointed out that our predicted partial width
ratio
\begin{eqnarray}
\frac{\Gamma[\chi_{c2}(2P)\to
\psi(2S)\gamma]}{\Gamma[\chi_{c2}(2P)\to J/\psi\gamma]}\simeq 4,
\end{eqnarray}
is in good agreement with that of GI and NR potential model, though
their predicted partial widths are about a factor of $2-3$ larger
than our results. Our predicted ratio is slightly larger than the
result $2.95\pm 0.5$ predicted with an effective Lagrangian
approach~\cite{DeFazio:2008xq}. To better understand the properties
of the $\chi_{c2}(2P)$ state and test the various theoretical
predictions, this ratio is suggested to be measured in experiments.

\subsubsection{$\chi_{c1}(2P)$}

The $X(3872)$ resonance has the same quantum numbers as
$\chi_{c1}(2P)$ (i.e., $J^{PC}=1^{++}$) and a similar mass to the
predicted value of $\chi_{c1}(2P)$. However, its exotic properties
can not be well understood with a pure $\chi_{c1}(2P)$
state~\cite{Olsen:2014qna,Voloshin:2007dx}. To understand the nature
of $X(3872)$, measurements of the radiative decays of $X(3872)$ have
been carried out by the BaBar~\cite{Aubert:2008ae},
Belle~\cite{Bhardwaj:2011dj}, and LHCb~\cite{Aaij:2014ala}
collaborations, respectively. Obvious evidence of $X(3872)\to J/\psi
\gamma$ was observed by these collaborations. Furthermore, the BaBar
and LHCb Collaborations also observed evidence of $X(3872)\to
\psi(2S) \gamma$. The branching fraction ratio
\begin{eqnarray}
R^{\mathrm{exp}}_{\psi'\gamma/\psi\gamma}=\frac{\Gamma[X(3872)\to
\psi(2S) \gamma]}{\Gamma[X(3872)\to J/\psi \gamma]}\simeq 3.4\pm
1.4,
\end{eqnarray}
obtained by the BaBar Collaboration~\cite{Aubert:2008ae} is
consistent with the recent measurement
$R^{\mathrm{exp}}_{\psi'\gamma/\psi\gamma}=2.46\pm 0.93$ of LHCb
Collaboration~\cite{Aaij:2014ala}.

Considering the $X(3872)$ as a pure $\chi_{c1}(2P)$ state, we
calculate the radiative decays $X(3872)\to J/\psi \gamma,\psi(2S)
\gamma$. Our predicted radiative decay partial widths are
\begin{eqnarray}
\Gamma[X(3872)\to J/\psi\gamma])& \simeq & 14.4 \
\mathrm{keV},\\
\Gamma[X(3872)\to \psi(2S)\gamma]& \simeq & 57.1 \ \mathrm{keV}.
\end{eqnarray}
It should be mentioned that the M2 transitions could give a $\sim
10\%$ correction to the radiative decay widths. Combing these
predicted partial widths, we can easily obtain the ratio
\begin{eqnarray}
R^{\mathrm{th}}_{\psi'\gamma/\psi\gamma}=\frac{\Gamma[X(3872)\to
\psi(2S) \gamma]}{\Gamma[X(3872)\to J/\psi \gamma]}\simeq 4.0,
\end{eqnarray}
which is consistent with the BaBar's
measurement~\cite{Aubert:2008ae}, and close to the upper limit of
the observations from LHCb~\cite{Aaij:2014ala}. The ratio predicted
by us is also in agreement with the result
$R_{\psi'\gamma/\psi\gamma}^{\mathrm{th}}\simeq 4.4$ from a
relativistic Salpeter method~\cite{Li:2009zu,Wang:2010ej}. It should
be mentioned that Barnes and Godfrey gave a ratio
$R_{\psi'\gamma/\psi\gamma}^{\mathrm{th}}\simeq
6$~\cite{Barnes:2003vb}, which is obviously larger than ours although
their predicted partial widths are close to ours.

Furthermore, considering $X(3872)$ as a pure $\chi_{c1}(2P)$ state,
we calculate the radiative decays into the $D$-wave states
$\psi(3770)$ and $\psi_2(1D)$ (i.e., $X(3823)$):
\begin{eqnarray}
\Gamma[X(3872)\to \psi(3770)\gamma]& \simeq & 1.8 \
\mathrm{keV},\\
\Gamma[X(3872)\to \psi_2(1D) \gamma]& \simeq & 0.77 \ \mathrm{keV}.
\end{eqnarray}
These partial widths are much smaller than the partial widths into
the $S$-wave states. Combining the measured width of $X(3872)$
(i.e., $\Gamma< 1.2$ MeV) from the PDG~\cite{PDG}, we estimate the
branching ratios
\begin{eqnarray}
\mathcal{B}[X(3872)\to \psi(3770)\gamma])& > & 1.9\times 10^{-3},\\
\mathcal{B}[X(3872)\to \psi_2(1D) \gamma]& > & 6.4 \times 10^{-4}.
\end{eqnarray}
The sizeable decay rates indicate that the decay modes
$\psi(3770)\gamma$ and $\psi_2(1D) \gamma$ might be observed in
experiments if $X(3872)$ corresponds to the $\chi_{c1}(2P)$ state
indeed.

As a whole, from the view of branching fraction ratio
$R_{\psi'\gamma/\psi\gamma}$, our result supports $X(3872)$ as a
candidate of $\chi_{c1}(2P)$, which is in agreement with the
predictions in Refs.~\cite{Li:2009zu,Wang:2010ej}. More observations
in the $\psi(3770)\gamma$ and $\psi_2(1D) \gamma$ channels are
useful to understand the nature of $X(3872)$.

\subsubsection{$\chi_{c0}(2P)$}

The $\chi_{c0}(2P)$ state is still not well-established, although
$X(3915)$ was recommended as the $\chi_{c0}(2P)$ state in
Ref.~\cite{Liu:2009fe}, and also assigned as the $\chi_{c0}(2P)$
state by the PDG~\cite{PDG} recently. Assigning the $X(3915)$ as the
$\chi_{c0}(2P)$ state will face serval serious
problems~\cite{Guo:2012tv,Olsen:2014maa}. For example, the mass
splitting between $\chi_{c0}(2P)$ and $\chi_{c2}(2P)$, the
production rates, and limits on the $\omega J/\psi$ and $D\bar{D}$
branching fractions of $X(3915)$ can not be well explained. Guo and
Meissner refitted the BaBar and Belle data of $\gamma\gamma \to
D\bar{D}$ separately, their analysis indicates that the broad bump
in the invariant mass spectrum could be assigned as the
$\chi_{c0}(2P)$ state~\cite{Guo:2012tv}. Its average mass and width
are $M=(3837.6\pm 11.5)$ MeV and $\Gamma=(221\pm 19)$ MeV,
respectively~\cite{Guo:2012tv}. The extracted mass of
$\chi_{c0}(2P)$ is consistent with the predictions in the screened
potential model~\cite{Li:2009zu} and relativistic quark
model~\cite{Ebert:2003}. Recently, Zhou \emph{et al.} carried out a
combined amplitude analysis of the $\gamma\gamma \to D\bar{D},
\omega J/\psi$ data~\cite{Zhou:2015uva}. They demonstrated that
$X(3915)$ and $X(3930)$ can be regarded as the same state with
$J^{PC}=2^{++}$ (i.e., $\chi_{c2}(2P)$). To establish the
$\chi_{c0}(2P)$ state and clarify the controversial situation of
$X(3915)$, a study of the radiative transitions of $\chi_{c0}(2P)$
in both theory and experiment is necessary.

\begin{figure}[ht]
\centering \epsfxsize=8.6 cm \epsfbox{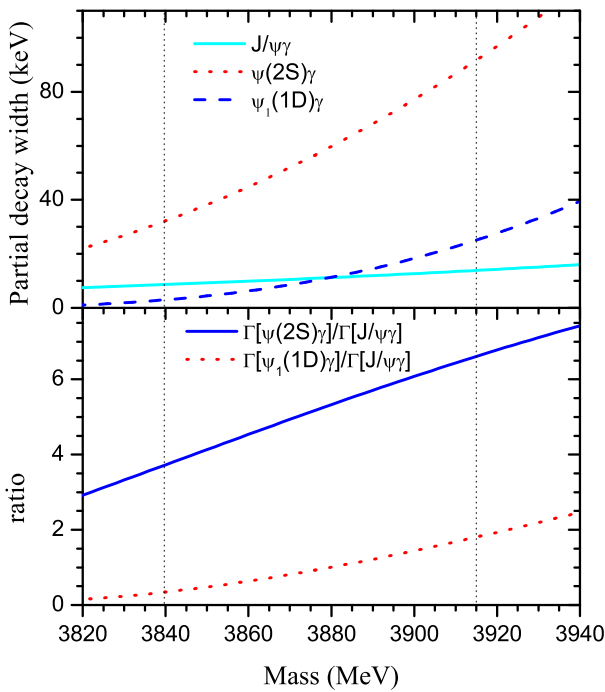} \caption{(Color
online) The partial widths of the radiative transitions of
$\chi_{c0}(2P)$, and their ratios as a function of the mass in the
range of $(3820\sim 3940)$ GeV. }\label{3p0}
\end{figure}

The $\chi_{c0}(2P)$ state can decay via the radiative transitions
$\chi_{c0}(2P)\to \psi(3770)\gamma, \psi(2S)\gamma, J/\psi \gamma$.
For we do not know the accurate mass of $\chi_{c0}(2P)$, we
calculate the properties of these radiative transitions as a
function of the mass in the region $(3820\sim 3940)$ GeV. Our
results are shown in Fig.~\ref{3p0}. From the figure, one can find
that the radiative transitions of $\chi_{c0}(2P)\to
\psi(3770)\gamma, \psi(2S)\gamma$ are sensitive to the mass of
$\chi_{c0}(2P)$.

If $\chi_{c0}(2P)$ has a low mass of $M\simeq3840$ MeV as suggested
by Guo and Meissner~\cite{Guo:2012tv}, $\psi(2S)\gamma$ and $J/\psi
\gamma$ are the important radiative decay modes. The predicted
partial decay widths are
\begin{eqnarray}
\Gamma[\chi_{c0}(2P)\to J/\psi\gamma]& \simeq & 8.6 \
\mathrm{keV},\\
\Gamma[\chi_{c0}(2P)\to \psi(2S)\gamma]& \simeq & 32 \
\mathrm{keV},\\
\Gamma[\chi_{c0}(2P)\to \psi(3770)\gamma]& \simeq & 2.1 \
\mathrm{keV},
\end{eqnarray}
and the predicted partial width ratios are
\begin{eqnarray}
\frac{\Gamma[\chi_{c0}(2P)\to
\psi(2S)\gamma]}{\Gamma[\chi_{c0}(2P)\to J/\psi\gamma]}& \simeq &
3.7,\\
\frac{\Gamma[\chi_{c0}(2P)\to
\psi(3770)\gamma]}{\Gamma[\chi_{c0}(2P)\to J/\psi\gamma]}& \simeq &
0.24.
\end{eqnarray}

On the other hand, if $\chi_{c0}(2P)$ has a high mass of
$M\simeq3915$ MeV as suggested by the PDG~\cite{PDG}, the radiative
decay mode $\psi(3770)\gamma$ also becomes important. Our predicted
partial decay widths are
\begin{eqnarray}
\Gamma[\chi_{c0}(2P)\to J/\psi\gamma]& \simeq & 14 \
\mathrm{keV},\\
\Gamma[\chi_{c0}(2P)\to \psi(2S)\gamma]& \simeq & 92 \
\mathrm{keV},\\
\Gamma[\chi_{c0}(2P)\to \psi(3770)\gamma]& \simeq & 21 \
\mathrm{keV}.
\end{eqnarray}
While the predicted partial width ratios are
\begin{eqnarray}
\frac{\Gamma[\chi_{c0}(2P)\to
\psi(2S)\gamma]}{\Gamma[\chi_{c0}(2P)\to J/\psi\gamma]}& \simeq &
6.6,\\
\frac{\Gamma[\chi_{c0}(2P)\to
\psi(3770)\gamma]}{\Gamma[\chi_{c0}(2P)\to J/\psi\gamma]}& \simeq &
1.5.
\end{eqnarray}

According to our calculations, we find that the properties of the
radiative transitions of $\chi_{c0}(2P)$ between the low- and
high-mass cases are very different. Thus, we suggest our
experimental colleagues observe the $\chi_{c0}(2P)$ state in the
$\psi(2S)\gamma, J/\psi \gamma,\psi(3770)\gamma$ decay channels and
measure these partial width ratios, which might provide us a good
chance to clarify the puzzles about the $\chi_{c0}(2P)$ state.


\begin{figure}[ht]
\centering \epsfxsize=8.6 cm \epsfbox{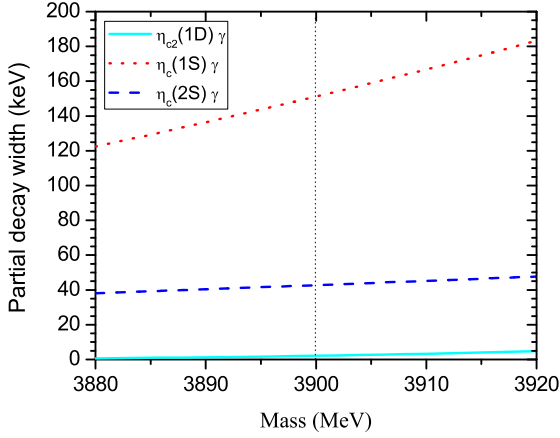} \caption{(Color
online) The partial decay widths of $h_{c}(2P)$ as a function of its
mass in the range of $M_{h_{c}(2P)}=(3900\pm 20)$ MeV.}\label{hc2p}
\end{figure}

\subsubsection{$h_{c}(2P)$}

There is no information of $h_{c}(2P)$ from experiments. According
to the model predictions, the mass-splitting between $\chi_{c2}(2P)$
and $h_{c}(2P)$ is about $M_{\chi_{c2}(2P)}-M_{h_{c}(2P)}=(30\pm
10)$ MeV~\cite{Swanson:2005,Li:2009zu,Cao:2012du}. Thus, the mass of
$h_{c}(2P)$ is most likely to be $M_{h_{c}(2P)}\simeq 3900$ MeV. The
typical radiative decay channels of $h_{c}(2P)$ are
$\eta_c(1S,2S)\gamma$ and $\eta_{c2}(1D) \gamma$.

Considering the uncertainties of the mass, in Fig.~\ref{hc2p} we
plot the partial decay widths as a function of mass in the range of
$M_{h_{c}(2P)}=(3900\pm 20)$ MeV. It is found that the radiative
transition rate of $h_{c}(2P)\to \eta_{c2}(1D) \gamma$ is very weak.
While $h_{c}(2P)$ has large transition rates to the
$\eta_c(1S)\gamma$ and $\eta_c(2S)\gamma$ channels. The
$h_{c}(2P)\to \eta_c(1S)\gamma$ transition shows some sensitivities
to the mass of $h_{c}(2P)$. With $M_{h_{c}(2P)}\simeq (3900\pm 20)$
MeV, our predicted partial widths are
\begin{eqnarray}
\Gamma(h_{c}(2P)\to \eta_c(1S) \gamma)& \simeq & 151\pm 30 \
\mathrm{keV},\\
\Gamma(h_{c}(2P)\to \eta_c(2S)\gamma)& \simeq & 43 \pm 5 \
\mathrm{keV}.
\end{eqnarray}
The rather sizeable partial widths of $h_{c}(2P)\to
\eta_c(1S,2S)\gamma$ are also obtained in the potential model
calculations~\cite{Swanson:2005,Li:2009zu,Cao:2012du}. Thus, the
radiative transitions $h_{c}(2P)\to \eta_c(1S,2S)\gamma$ are worth
to observing in experiments.

Recently, the BESIII Collaboration observed a new neutral
charmonium-like particle $Z_c(3900)^0$ in the $e^+e^-\to
\pi^0\pi^0J/\psi$ process~\cite{Ablikim:2015tbp}, which is likely
the isospin partner of $Z_c(3900)^{\pm}$. It should be mentioned
that the possibility of the $Z_c(3900)^0$ as a candidate of
$h_{c}(2P)$ should be considered carefully as well.

\subsection{Radiative transitions of $3S$ states}

\subsubsection{$\psi(4040)$}

The $\psi(4040)$ resonance is commonly identified with the
$\psi(3S)$ state~\cite{Eichten:2007qx}. This state can decay into
$\chi_{cJ}(1P)\gamma $ and $\chi_{cJ}(2P)\gamma $ via the radiative
transitions. We have calculated these precesses. According to our
calculations, it is found that the radiative transition rates of
$\psi(4040)\to \chi_{cJ}(1P)\gamma $ are relatively weak. Our
predicted partial widths are
\begin{eqnarray}
\Gamma[\psi(4040) \to \chi_{c0}(1P)\gamma]& \simeq & 3.1\ \mathrm{keV},\\
\Gamma[\psi(4040) \to \chi_{c1}(1P)\gamma]& \simeq & 3.0\ \mathrm{keV},\\
\Gamma[\psi(4040) \to \chi_{c2}(1P)\gamma]& \simeq & 2.2\
\mathrm{keV}.
\end{eqnarray}
The small partial decay widths of $\Gamma[\psi(4040) \to
\chi_{cJ}(1P)\gamma]$ are also predicted with potential
models~\cite{Swanson:2005,Cao:2012du}. Recently, the Belle
Collaboration observed the processes $e^+e^-\to
\chi_{c1,c2}(1P)\gamma$ \cite{Han:2015vhc}, they only gave upper
limits on branching fractions $\mathcal{B}[\psi(4040) \to
\chi_{c1}(1P)\gamma]= 3.4\times 10^{-3}$ and $\mathcal{B}[\psi(4040)
\to \chi_{c2}(1P)\gamma]= 5.5\times 10^{-3}$. Combing the measured
decay width, one obtains the partial decay widths of
$\Gamma[\psi(4040) \to \chi_{c1}\gamma]< 272$ keV and
$\Gamma[\psi(4040) \to \chi_{c2}\gamma]< 440$ keV from experiments.
The upper limits of these partial widths from experiments are too
large to give any constrains of the theoretical predictions. It might be
a challenge to carry out accurate observations of the radiative
transitions $\psi(4040)\to \chi_{cJ}(1P)\gamma $ in experiments for
their small decay rates.

\begin{figure}[ht]
\centering \epsfxsize=8.6 cm \epsfbox{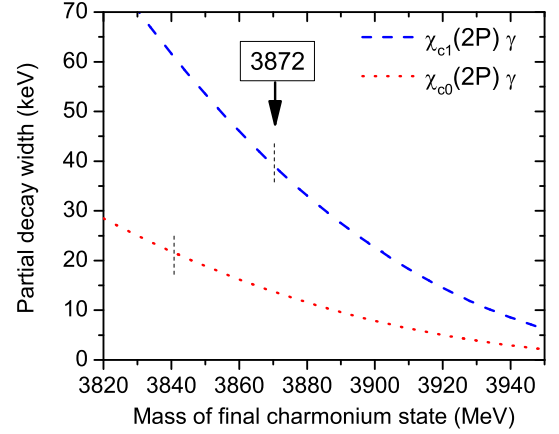} \caption{(Color
online) The partial decay widths of the radiative transitions
$\psi(4040)\to \chi_{c0}(2P)\gamma, \chi_{c1}(2P)\gamma$ as
functions of the masses of the final states $\chi_{c0}(2P)$ and
$\chi_{c1}(2P)$. }\label{ps3s}
\end{figure}

It is interesting to find that the radiative transition rates of
$\psi(4040)\to \chi_{cJ}(2P)\gamma $ are not small in theory. The
decay rates to the $\chi_{cJ}(2P)\gamma $ channels are about one
order of magnitude larger than those of the $\chi_{cJ}(1P)\gamma $
channels. We predict
\begin{eqnarray}
\Gamma[\psi(4040) \to \chi_{c2}(2P)\gamma]& \simeq & 19\
\mathrm{keV}.
\end{eqnarray}
Our result is compatible with the NR potential model
prediction~\cite{Swanson:2005}. The $\chi_{c2}(2P)\gamma$ decay mode
of $\psi(4040)$ is likely to be observed in forthcoming experiments.

The $\chi_{c0}(2P)$ and $\chi_{c1}(2P)$ states are still not
established, thus, we calculate the partial decay widths of the
radiative transitions $\psi(4040)\to \chi_{c0}(2P)\gamma,
\chi_{c1}(2P)\gamma$ as functions of the masses of the
$\chi_{c0}(2P)$ and $\chi_{c1}(2P)$ states in a possible region. Our
results are shown in Fig.~\ref{ps3s}. If the $X(3872)$ is the
$\chi_{c1}(2P)$ state, our predicted partial decay width of the
process $\psi(4040) \to \chi_{c1}(2P)\gamma$ is
\begin{eqnarray}
\Gamma[\psi(4040) \to \chi_{c1}(2P)\gamma]& \simeq & 38\
\mathrm{keV}.
\end{eqnarray}
While if the mass of $\chi_{c0}(2P)$ is $\sim 3840$ MeV as suggested
in Refs.~\cite{Li:2009zu,Guo:2012tv}, the predicted partial decay
width of the process $\psi(4040) \to \chi_{c0}(2P)\gamma$ is
\begin{eqnarray}
\Gamma[\psi(4040) \to \chi_{c0}(2P)\gamma]& \simeq & 22\
\mathrm{keV}.
\end{eqnarray}
Thus, if $X(3872)$ is identified as $\chi_{c1}(2P)$ indeed, and
$\chi_{c0}(2P)$ has a low mass of $\sim 3840$ MeV, both
$\chi_{c1}(2P)$ and $\chi_{c0}(2P)$ could be produced through the
radiative decays of $\psi(4040)$, and established in the
$J/\psi\gamma$ and $\psi(2S)\gamma$ final states.

\subsubsection{$\eta_c(3S)$}

The $\eta_c(3S)$ state is not established in experiments. According
to model predictions, its mass is about $40-100$ MeV lower than that
of $\psi(4040)$~\cite{Swanson:2005,Li:2009zu}. Thus, its mass is
most likely to be in the range of $3940- 4000$ MeV. The $X(3940)$
resonance observed by the Belle Collaboration in $e^+e^-\to J/\psi
X$~\cite{Abe:2007jna,Abe:2007sya} is a good candidate of
$\eta_c(3S)$ by comparing the observed decay mode, width, and mass
with those predicted in theory~\cite{Eichten:2007qx}.

Considering $X(3940)$ as the $\eta_c(3S)$ state, we calculate the
radiative transitions $\eta_c(3S)\to h_c(1P)\gamma, h_c(2P)\gamma$.
Our predicted partial decay widths
\begin{eqnarray}
\Gamma[\eta_c(3940) \to h_{c}(1P)\gamma]& \simeq & 1.8\
\mathrm{keV},\\
\Gamma[\eta_c(3940) \to h_{c}(2P)\gamma]& \simeq & 1.7\
\mathrm{keV},
\end{eqnarray}
are relatively small. Combining the observed width of $X(3940)$,
i.e., $\Gamma \simeq 37$ MeV, we easily predict the branching
fractions
\begin{eqnarray}
\mathcal{B}[\eta_c(3940) \to h_{c}(1P)\gamma]& \simeq & 4.9\times
10^{-5},\\
\mathcal{B}[\eta_c(3940) \to h_{c}(2P)\gamma]& \simeq & 4.6\times
10^{-5}.
\end{eqnarray}
The small branching ratios indicate that if $X(3940)$ is the
$\eta_c(3S)$ state, it might be difficult to be observed in the
radiative decay channels $h_{c}(1P)\gamma$ and $h_{c}(2P)\gamma$.
Note that we have taken the mass of $h_{c}(2P)$ with
$M_{h_{c}(2P)}=3.9$ GeV in the calculation. Finally, it should be
mentioned that our predicted partial widths are notably smaller than those of NR
and GI potential models~\cite{Swanson:2005} (see Tab.~\ref{EMPA}).

\subsection{Radiative transitions of $2D$ states}

\subsubsection{$\psi(4160)$}

The $1^{--}$ state $\psi(4160)$ is commonly identified with the $2^3
D_1$ state. The average experimental mass and width from the PDG are
$M=4191\pm 5$ MeV and $\Gamma=70\pm 10$ MeV,
respectively~\cite{PDG}. The $\psi(4160)$ resonance can decay into
$\chi_{cJ}(1P)\gamma$ and $\chi_{cJ}(2P)\gamma$ via the radiative
transitions.

Considering $\psi(4160)$ as a pure $2^3 D_1$ state, we predict
\begin{eqnarray}
\Gamma[\psi(4160) \to \chi_{c0}(1P)\gamma]& \simeq & 21 \ \mathrm{keV},\\
\Gamma[\psi(4160) \to \chi_{c1}(1P)\gamma]& \simeq & 6.5\ \mathrm{keV},\\
\Gamma[\psi(4160) \to \chi_{c2}(1P)\gamma]& \simeq & 8.8\
\mathrm{keV}.
\end{eqnarray}
Our predictions of $\Gamma[\psi(4160) \to \chi_{c0,1}(1P)\gamma]$
are close to those of potential models~\cite{Swanson:2005}, however,
our prediction of $\Gamma[\psi(4160) \to \chi_{c2}(1P)\gamma]$ is
about one order of magnitude larger than that of potential
models~\cite{Swanson:2005} (see Tab.~\ref{D-wave}). It should be
mentioned that the M2 transitions could give a $10-30\%$ correction
to the radiative decay widths by interfering with the E$1$ transitions.
Combining the measured decay width of
$\psi(4160)$ with our predicted partial widths, we estimate the branching fractions:
\begin{eqnarray}
\mathcal{B}[\psi(4160) \to \chi_{c0}(1P)\gamma]& \simeq & 3.0\times
10^{-4},\\
\mathcal{B}[\psi(4160) \to \chi_{c1}(1P)\gamma]& \simeq & 1.0\times
10^{-4},\\
\mathcal{B}[\psi(4160) \to \chi_{c2}(1P)\gamma]& \simeq & 1.3\times
10^{-4}.
\end{eqnarray}
The CLEO Collaboration only gave upper limits of
$\mathcal{B}[\psi(4160) \to \chi_{c1}(1P)\gamma]<7\times 10^{-3}$
and $\mathcal{B}[\psi(4160) \to \chi_{c2}(1P)\gamma]<13\times
10^{-3}$~\cite{Coan:2006rv}. Our predictions are in the region of
the measurements. Of course, we expect that more accurate
observations can be carried out in future experiments.

Furthermore, we calculate the partial decay width of
$\Gamma[\psi(4160) \to \chi_{c2}(2P)\gamma]$. Our result
\begin{eqnarray}
\Gamma[\psi(4160) \to \chi_{c2}(2P)\gamma]& \simeq & 5.9\
\mathrm{keV},
\end{eqnarray}
is in agreement with the predictions in various potential
models~\cite{Swanson:2005,Li:2012vc}. At the same time, we can
easily estimate the branching fraction
\begin{eqnarray}
\mathcal{B}[\psi(4160) \to \chi_{c2}(2P)\gamma] \simeq 8.1\times
10^{-5}.
\end{eqnarray}
The small branching fraction indicates that the
$\chi_{c2}(2P)\gamma$ decay mode might be difficult to be observed
in experiments.

\begin{figure}[ht]
\centering \epsfxsize=8.6 cm \epsfbox{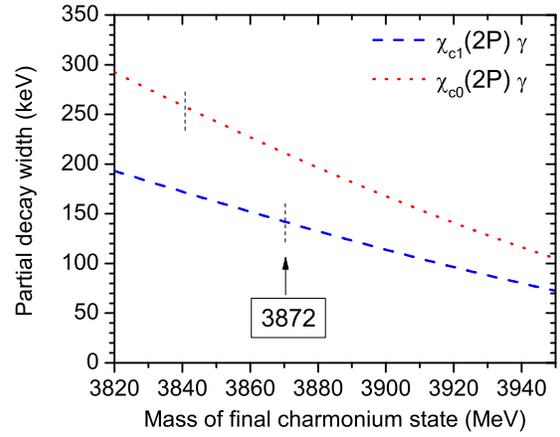} \caption{(Color
online) The partial decay widths of the radiative transitions
$\psi(4160)\to \chi_{c0}(2P)\gamma, \chi_{c1}(2P)\gamma$ as
functions of the masses of the final states $\chi_{c0}(2P)$ and
$\chi_{c1}(2P)$.}\label{3d1}
\end{figure}

The $\chi_{c0}(2P)$ and $\chi_{c1}(2P)$ states are still not
established, thus, we calculate the partial decay widths of the
radiative transitions $\psi(4160)\to \chi_{c0}(2P)\gamma,
\chi_{c1}(2P)\gamma$ as functions of the masses of the
$\chi_{c0}(2P)$ and $\chi_{c1}(2P)$ states in a possible region. Our
results are shown in Fig.~\ref{3d1}. From the figure, it is seen
that although the partial decay widths of $\Gamma[\psi(4160)\to
\chi_{c0}(2P)\gamma, \chi_{c1}(2P)\gamma]$ are sensitive to the mass
of the final charmonium states, the partial decay widths are still
fairly large when we take the upper limits for the masses of the
final charmonium states. Thus, $\psi(4160)$ might be a good source
to be used to look for the missing $\chi_{c0}(2P)$ and
$\chi_{c1}(2P)$ states via the radiative transitions $\psi(4160)\to
\chi_{c0}(2P)\gamma, \chi_{c1}(2P)\gamma$, which is also suggested
by Li, Meng and Chao in Ref.~\cite{Li:2012vc}.

Considering $X(3872)$ as the $\chi_{c1}(2P)$ state, we predict
partial decay width and branching fraction of the process
$\psi(4160) \to \chi_{c1}(2P)\gamma$:
\begin{eqnarray}
\Gamma(\psi(4160) \to \chi_{c1}(2P)\gamma)& \simeq & 140\
\mathrm{keV},\\
\mathcal{B}[\psi(4160) \to \chi_{c1}(2P)\gamma]& \simeq & 2.0\times
10^{-3}.
\end{eqnarray}
While if the mass of $\chi_{c0}(2P)$ is $\sim 3840$ MeV as predicted
in Refs.~\cite{Li:2009zu,Guo:2012tv}, the predicted partial decay
width and branching fraction of the process $\psi(4160) \to
\chi_{c0}(2P)\gamma$ are
\begin{eqnarray}
\Gamma(\psi(4160) \to \chi_{c0}(2P)\gamma)& \simeq & 259\
\mathrm{keV},\\
\mathcal{B}[\psi(4160) \to \chi_{c0}(2P)\gamma]& \simeq & 8.6\times
10^{-3}.
\end{eqnarray}
According to our calculations, if $X(3872)$ is $\chi_{c1}(2P)$ and
the mass of $\chi_{c0}(2P)$ is $\sim 3840$ MeV indeed, these states
should be easily observed in the radiative transitions
$\psi(4160)\to \chi_{c0}(2P)\gamma, \chi_{c1}(2P)\gamma$ for their
large branching fractions. To clarify the puzzles in $X(3872)$ and
$\chi_{c0}(2P)$, we strongly suggest our experimental colleagues
carry out observations of the radiative transitions $\psi(4160)\to
\chi_{c0}(2P)\gamma, \chi_{c1}(2P)\gamma$ in the future.

\subsubsection{$\psi_2(2D)$}

The $\psi_2(2D)$ state is still not observed in experiments.
Combined the measured mass of $\psi_1(2D)$ and the $2^3D_2-2^3D_1$
mass splitting from the quark model~\cite{Swanson:2005}, the mass of
$\psi_2(2D)$ may be about $M_{\psi_2(2D)}\simeq 4208$ MeV. The study
of the radiative transitions $\psi_2(2D)\to \chi_{cJ}(1P,2P)\
\gamma$ may be helpful to look for $\psi_2(2D)$ in experiments.

According to our calculations, we find that the partial widths of
$\Gamma[\psi_2(2D)\to \chi_{c0}(1P,2P)\gamma]$ are negligibly small.
The $\chi_{c1}(1P)$, $\chi_{c2}(1P)$ and $\chi_{c2}(2P)$ states are
well-established, thus, we can easily predict that
\begin{eqnarray}
\Gamma[\psi_2(2D) \to \chi_{c1}(1P)\gamma]& \simeq & 26 \ \mathrm{keV},\\
\Gamma[\psi_2(2D)\to \chi_{c2}(1P)\gamma]& \simeq & 10 \
\mathrm{keV},\\
\Gamma[\psi_2(2D) \to \chi_{c2}(2P)\gamma]& \simeq & 64 \
\mathrm{keV},
\end{eqnarray}
which are close to the predictions of NR potential
model~\cite{Swanson:2005}. It should be mentioned that the
E3 together with the M2 transitions have obvious corrections to
the partial widths by interfering with E1 transitions (see Tab.~\ref{D-wave}).

\begin{figure}[ht]
\centering \epsfxsize=8.6 cm \epsfbox{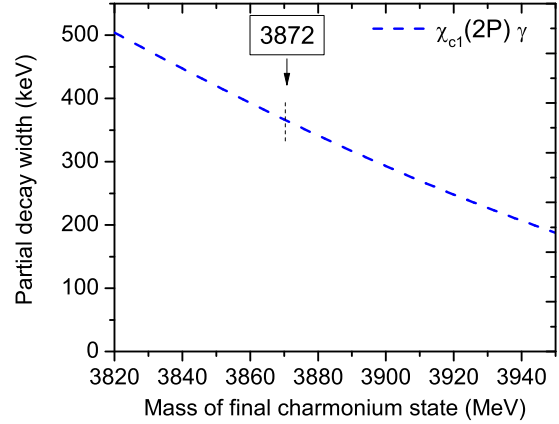} \caption{(Color
online) The partial decay widths of the radiative transitions
$\psi_2(2D)\to \chi_{c1}(2P)\gamma$ as a function of the mass of the
final state $\chi_{c1}(2P)$.}\label{3d2}
\end{figure}

The $\chi_{c1}(2P)$ state is still not established, thus,
we calculate the partial decay width of the radiative transition
$\psi_2(2D)\to \chi_{c1}(2P)\gamma$ as a function of the mass of the
$\chi_{c1}(2P)$ state in a possible region. Our results are shown in
Fig.~\ref{3d2}. From the figure, it is seen that $\psi_2(2D)$ has a
large transition rate into $\chi_{c1}(2P)\gamma$. The radiative
partial width of $\Gamma[\psi_2(2D)\to \chi_{c1}(2P)\gamma]$ is
about 100s keV, which is in agreement with the potential model
predictions~\cite{Swanson:2005}. If $X(3872)$ is assigned as the
$\chi_{c1}(2P)$ state, we predict that
\begin{eqnarray}
\frac{\Gamma[\psi_2(2D) \to \chi_{c1}(2P)\gamma]}{\Gamma[\psi_2(2D)
\to \chi_{c2}(2P)\gamma]}& \simeq & 5.6.
\end{eqnarray}
This ratio might be interesting to test the nature of $X(3872)$ in
future experiments.

\subsubsection{$\psi_3(2D)$}

The $\psi_3(2D)$ state is still not established. Combined the
measured mass of $\psi_1(2D)$ and the $2^3D_3-2^3D_1$ mass splitting
from the quark model~\cite{Swanson:2005}, the mass of $\psi_3(2D)$
is about $M_{\psi_3(2D)}\simeq 4217$ MeV. The $\psi_3(2D)$ can decay
into $\chi_{cJ}(1P)\gamma$ and $\chi_{cJ}(2P)\gamma$ via radiative
transitions.

We find that $\chi_{c2}(1P)\gamma$ and $\chi_{c2}(2P)\gamma$ are the
dominant radiative decay channels of $\psi_3(2D)$. We predict their
partial decay widths:
\begin{eqnarray}
\Gamma[\psi_3(2D) \to \chi_{c2}(1P)\gamma]& \simeq & 25 \ \mathrm{keV},\\
\Gamma[\psi_3(2D) \to \chi_{c2}(2P)\gamma]& \simeq & 349 \
\mathrm{keV},
\end{eqnarray}
which are compatible with the potential model
results~\cite{Swanson:2005}. The large partial width of $\Gamma
[\psi_3(2D)\to \chi_{c2}(2P)\gamma]$ indicates that
$\chi_{c2}(2P)\gamma$ might be a good channel to find the missing
state $\psi_3(2D)$ in future experiments.

Furthermore, in our calculations we find that the partial decay
widths of $\Gamma[\psi_3(2D)\to \chi_{c0,c1}(2P)\gamma]$ are
negligibly small, while the $\Gamma[\psi(2^3D_3)\to
\chi_{c0,c1}(1P)\gamma]$ are sizeable. Our results are
\begin{eqnarray}
\Gamma[\psi_3(2D) \to \chi_{c0}(1P)\gamma]& \simeq & 8.5 \ \mathrm{keV},\\
\Gamma[\psi_3(2D) \to \chi_{c1}(1P)\gamma]& \simeq & 9.1 \
\mathrm{keV}.
\end{eqnarray}
It should be pointed out that in these two decay processes the
E1 translations are nearly forbidden. Their partial widths
are mainly contributed by the E3 translations.

\subsubsection{$\eta_{c2}(2D)$}

There is no information of $\eta_{c2}(2D)$ from experiments.
Combined the measured mass of $\psi_1(2D)$ and the $2^1D_2-2^3D_1$
mass splitting from the quark model~\cite{Swanson:2005}, the mass of
$\eta_{c2}(2D)$ is $M_{\eta_{c2}(2D)}\simeq 4208$ MeV. We study the
typical radiative transitions $\eta_{c2}(2D)\to h_{c}(1P,2P)\gamma$.
We find that $\eta_{c2}(2D)$ might have strong decay rates into both
$h_{c}(1P)\gamma$ and $h_{c}(2P)\gamma$ channels. We predict that
\begin{eqnarray}
\Gamma[\eta_{c2}(2D) \to h_{c}(1P)\gamma]& \simeq & 39 \ \mathrm{keV},\\
\Gamma[\eta_{c2}(2D) \to h_{c}(2P)\gamma]& \simeq & 385 \
\mathrm{keV},
\end{eqnarray}
where we take the mass of $h_{c}(2P)$ with $M_{h_{c}(2P)}=3900$ MeV.
Our predictions are in agreement with those from potential
models~\cite{Swanson:2005} in magnitude (see Tab.~\ref{D-wave}).
According to our analysis, the observations of the decay chain:
$\eta_{c2}(2D) \to h_{c}(2P)\gamma$, $h_{c}(2P)\to \eta_c(1S)\gamma$
might be useful for the search for $\eta_{c2}(2D)$ and $h_{c}(2P)$
in experiments.

\begin{figure*}[ht]
\centering \epsfxsize=17.2 cm \epsfbox{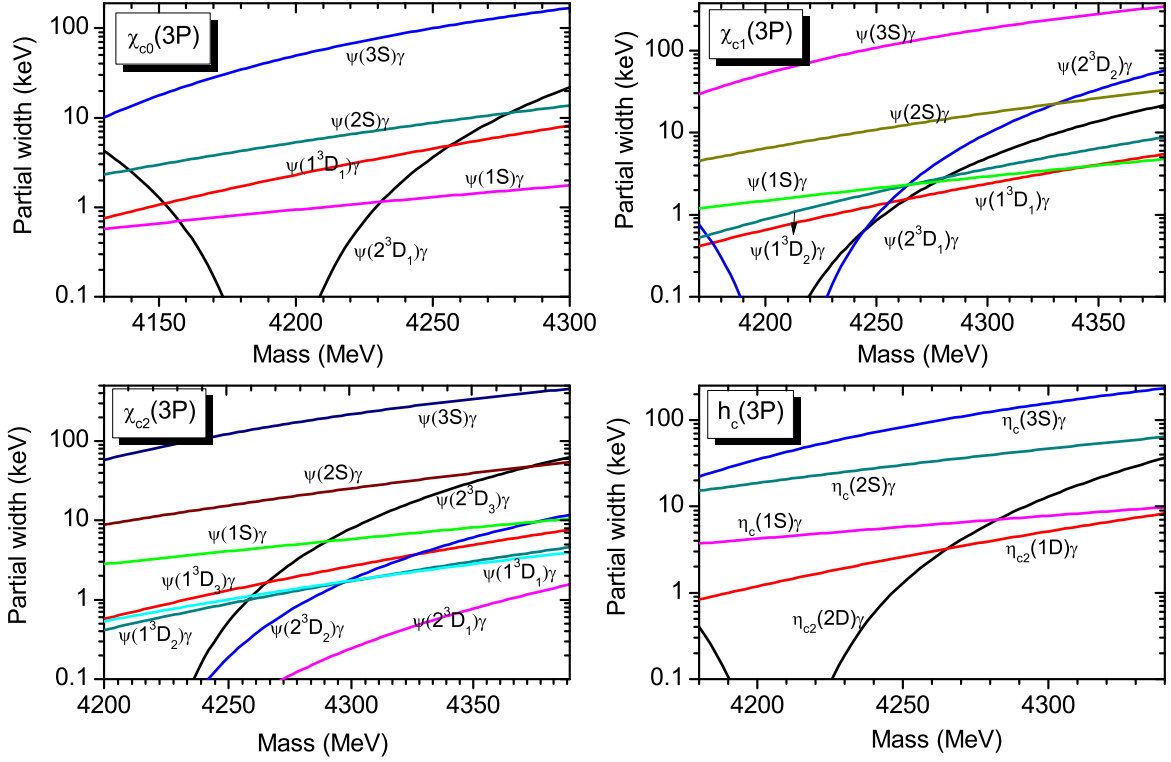} \caption{(Color
online) The partial radiative decay width (keV) of the radiative
transitions of $3P$ states as a function of mass.}\label{cc3p}
\end{figure*}

\subsection{Radiative transitions of $3P$ states}

Until now, no $3P$ charmonium states, $\chi_{c0,1,2}(3P)$ and
$h_{c}(3P)$, have been established in experiments. Their predicted
masses from various quark models have large uncertainties. The
possible masses of the $\chi_{c0}(3P)$, $\chi_{c1}(3P)$,
$\chi_{c2}(3P)$, and $h_{c}(3P)$ are in the ranges of ($4.13-4.30$)
GeV, ($4.17-4.38$) GeV, ($4.20-4.39$) GeV, and ($4.18-4.34$) GeV,
respectively~\cite{Li:2009zu,Swanson:2005,Cao:2012du}. Considering
the uncertainties of their predicted masses, we plot the partial
radiative decay width as a function of mass in Fig.~\ref{cc3p}. From
the figure, one can see that the radiative decay properties of the
$3P$ charmonium states are very sensitive to their masses. The EM
decays of $\chi_{cJ}(3P)$ and $h_{c}(3P)$ are governed by the
$\psi(3S)\gamma$ and $\eta(3S)\gamma$ channels, respectively. The
partial widths of the $3P$ charmonium states decaying into the $2S$
states are also sizeable. If the $3P$ charmonium states have a
larger mass, their partial width decaying into the $2D$ states will
become comparable to those of into the $2S$ states.


It is interesting to note that the new $X(4350)$ state observed by
the Belle Collaboration \cite{Shen:2009vs} was recommended as the
$\chi_{c2}(3P)$ state by Liu \emph{et al.}~\cite{Liu:2009fe}. If
$X(4350)$ corresponds to $\chi_{c2}(3P)$, its partial decay widths
decaying into the $\psi(1S,2S,3S)\gamma$ channels are
\begin{eqnarray}
\Gamma[X(4350) \to \psi(3S)\gamma]& \simeq & 341 \ \mathrm{keV},\\
\Gamma[X(4350) \to \psi(2S)\gamma]& \simeq & 39 \ \mathrm{keV},\\
\Gamma[X(4350) \to \psi(1S)\gamma]& \simeq & 8 \ \mathrm{keV}.
\end{eqnarray}
Combining these predicted partial widths with the measured width $\Gamma_{\mathrm{exp}}\simeq 13$ MeV
of $X(4350)$, we can estimate the branching fractions:
\begin{eqnarray}
\mathcal{B}[X(4350) \to \psi(3S)\gamma]& \simeq & 2.6\times
10^{-2},\\
\mathcal{B}[X(4350) \to \psi(2S)\gamma]& \simeq & 2.2\times
10^{-3},\\
\mathcal{B}[X(4350) \to \psi(1S)\gamma]& \simeq & 6.2\times 10^{-4}.
\end{eqnarray}
The large branching fractions of $\mathcal{B}[X(4350) \to
\psi(2S,3S)\gamma]$ indicate that $X(4350)$ should be observed in
the $\psi(3S)\gamma$ and $\psi(2S)\gamma$ channels. Furthermore, as
a candidate of $\chi_{c2}(3P)$ the radiative transition rate of
$X(4350) \to \psi_3(2D)\gamma$ should be sizeable. Our predicted
partial width and branching fraction are
\begin{eqnarray}
\Gamma[X(4350) \to \psi_3(2D)\gamma]& \simeq & 31 \ \mathrm{keV},\\
\mathcal{B}[X(4350) \to \psi_3(2D)\gamma]& \simeq & 2.4\times
10^{-3}.
\end{eqnarray}
Thus, if the $X(4350)$ state is the $\chi_{c2}(3P)$ state indeed,
$X(4350)$ might provide us a good source to look for the missing
state $\psi(2^3D_3)$ via its radiative transition.



If $X(4350)$ is identified as the $\chi_{c2}(3P)$ state, its mass is
very close to the predictions of GI potential model, thus, the
masses of $\chi_{c0}(3P)$, $\chi_{c1}(3P)$ and $h_{c}(3P)$ might be
$\sim 4.33$ GeV~\cite{Swanson:2005}. To compare our predictions of
the radiative transition properties of the $3P$ states with the
potential models', in Tab.~\ref{EMPA} we further list our results
predicted with the masses from GI, NR~\cite{Swanson:2005}, and
SNR~\cite{Li:2009zu} potential models, respectively. It is found
that the dominant decay channels predicted by us are consistent with
those of potential models, although our predicted values show some
obvious differences from those models'.

\section{Summary}\label{sum}

In the constituent quark model framework, we systematically study
the EM transitions of $nS,nP$ ($n\leq 3$), and $nD$ ($n\leq 2$)
charmonium states. Without introducing any free parameters, we
obtain a reasonable description of the EM transitions of the
well-established low-lying charmonium states $J/\psi$, $\psi(2S)$,
$\chi_{cJ}(1P)$, $h_c(1P)$ and $\psi(3770)$. It should be emphasized
that the M2 transitions might give obvious corrections to some E1 dominant processes by
interfering with the E1 transitions. We summarize our major
results as follows.

For the radiative transition $J/\psi\rightarrow \eta_c(1S) \gamma$,
there still exist puzzles in both theory and experiments.  Our
prediction of $\Gamma[J/\psi\rightarrow \eta_c(1S) \gamma]\simeq
1.25$ keV is consistent with the average experimental data from the
PDG~\cite{PDG}, and the calculations with a relativistic quark
model~\cite{Ebert:2003}. However, the calculations from potential
models~\cite{Swanson:2005} and a recent lattice QCD
approach~\cite{Becirevic:2012dc} gave a large partial decay width,
which is consistent with the recent measurements at KEDR
$\Gamma[J/\psi\rightarrow \eta_c(1S) \gamma]\simeq 2.98$
keV~\cite{Anashin:2014wva}. To clarify the discrepancies between
different experiments and test various model predictions, more
accurate observations of the $J/\psi\rightarrow \eta_c(1S) \gamma$
transition are needed in future experiments.

For the radiative transitions of $2S$ states, only some measurements
for $\psi(2S)$ can be obtained until now. About the radiative
transitions of $\eta_c(2S)$ it is found that our predictions have
notable differences from the other predictions. While, our
predictions about $\psi(2S)$ are comparable to the experimental
observations and other model predictions.
The partial width ratios for $\psi(2S)$, $\frac{\Gamma[\psi(2S)\to
\chi_{c1,2}(1P)\gamma]}{\Gamma[\psi(2S)\to \chi_{c0}(1P)\gamma]}$
and $\frac{\Gamma[\psi(2S)\to \eta_{c}(1S)\gamma]}{\Gamma[J/\psi\to
\eta_{c}(1S)\gamma]}$, have a strong model dependence. These partial
width ratios for $\psi(2S)$, and the radiative transitions of
$\eta_c(2S)\to h_c(1P)\gamma,J/\psi \gamma$ are worth to measuring
to test various theoretical approaches.

For the radiative transitions of $1P$ charmonium states, our
predictions are in reasonable agreement with the observations.
There are large discrepancies between different model predictions of
$\Gamma [h_{c}(1P)\to \eta_c \gamma]$, however, the present world
data can not be used to test the various predictions for their large
uncertainties. We hope more accurate measurements of the total decay
widths of $\chi_{c2}(1P)$ and $h_{c}(1P)$, together with the
observations of the $\chi_{c2}(1P)\to J/\psi \gamma$ and
$h_{c}(1P)\to \eta_c \gamma$ processes can be carried out in future
experiments.

For the radiative transitions of $1D$ charmonium states $\psi(3770)$
and $X(3823)$, our predictions are consistent with the observations.
For the missing $1D$ charmonium states $\eta_{c2}(1D)$ and
$\psi_{3}(1D)$, they have large radiative transition rates into
$h_c(1P)\gamma$ and $\chi_{c2}(1P)\gamma$, respectively. Their
corresponding partial decay widths might be $200-300$ keV, which
indicates that the missing states $\eta_{c2}(1D)$ and $\psi_{3}(1D)$
are most likely to be established in the $\eta_{c2}(1D)\to
h_c(1P)\gamma$ and $\psi_{3}(1D)\to\chi_{c2}(1P)\gamma$ processes,
respectively.

The main radiative decay modes for the triplet $2P$ states $\chi_{cJ}(2P)$ are
$\psi(2S)\gamma$ and $J/\psi(1S)\gamma$, while for the singlet $2P$
state $h_c(2P)$ are $\eta_c(1S,2S)\gamma$. The partial decay widths
of $\Gamma[\chi_{cJ}(2P)\to \psi(3770)\gamma]$ are also sizeable.
Considering the $X(3872)$ resonance as the $\chi_{c1}(2P)$ state, we
predict its radiative transition properties. Our predicted branching
fraction ratio $R_{\psi'\gamma/\psi\gamma}$ supports $X(3872)$ as a
candidate of $\chi_{c1}(2P)$. Further observations in the
$\psi(3770)\gamma$ and $\psi_2(1D) \gamma$ channels are necessary to
understand the nature of $X(3872)$. For the $\chi_{c0}(2P)$ state,
we study its radiative transition properties by setting its mass
with 3840 MeV and 3915 MeV, respectively. We find that the partial
width ratio $\frac{\Gamma[\chi_{c0}(2P)\to
\psi(2S)\gamma]}{\Gamma[\chi_{c0}(2P)\to J/\psi\gamma]}$ is
sensitive to the mass of $\chi_{c0}(2P)$. Thus, we suggest the
experimental Collaborations carry out some measurements of this
ratio to establish the $\chi_{c0}(2P)$ state in experiments finally.
For the missing state $h_{c}(2P)$, the predicted mass is
$M_{h_{c}(2P)}\simeq 3900$ MeV. The partial decay widths into the
$\eta_c(1S)\gamma$ and $ \eta_c(2S)\gamma$ channels are rather
sizeable. The discovery of the missing $h_{c}(2P)$ state in the
$\eta_c(1S)\gamma$ and $ \eta_c(2S)\gamma$ channels might be
possible in future experiments.

For the radiative transitions of the $\psi(3S)$ state, i.e.
$\psi(4040)$, we predict that the partial decay widths of
$\Gamma[\psi(3S)\to \chi_{cJ}(1P)\gamma] $ is only $2-3$ keV. It
might be a challenge to carry out accurate observations of the
radiative transitions $\psi(4040)\to \chi_{cJ}(1P)\gamma $ in
experiments for their small decay rates. However, the radiative
transition rate of $\psi(4040)\to \chi_{cJ}(2P)\gamma $ is fairly
large, which is about an order of magnitude larger than that of the
$\psi(4040)\to \chi_{cJ}(1P)\gamma $. The missing states
$\chi_{c1}(2P)$ and $\chi_{c0}(2P)$ may be produced through
radiative decays of $\psi(4040)$, and be established in the
$J/\psi\gamma$ and $\psi(2S)\gamma$ final states. Considering the
$X(3940)$ as a candidate of the singlet $\eta_c(3S)$ state, we
predict the radiative transition rates of $\eta_c(3S)\to
h_c(1P,2P)\gamma$ are small. Their branching ratios are about
$10^{-5}$, which indicates that if $X(3940)$ is the $\eta_c(3S)$
state, it might be difficult to be observed in its radiative
transitions.

Taking the $\psi(4160)$ as a pure $\psi_1(2D)$ state, we calculate
its radiative transition properties. We predict that the radiative
transition rates of $\psi(4160)$ into the $2P$ states
$\chi_{c0,1}(2P)$ are fairly large, the branching ratios of
$\mathcal{B}[\psi(4160)\to \chi_{c0,1}(2P) \gamma]$ are about
$10^{-2}-10^{-3}$. Thus, $\psi(4160)$ might be a very good source to be
used to look for the missing $\chi_{c0}(2P)$ and $\chi_{c1}(2P)$
states via the radiative transitions $\psi(4160)\to
\chi_{c0}(2P)\gamma, \chi_{c1}(2P)\gamma$. For the missing $2D$
states $\psi_2(2D)$, $\psi_3(2D)$ and $\eta_{c2}(2D)$, we find that
their partial widths of $\Gamma[\psi_2(2D)\to \chi_{c1}(2P)\gamma]$,
$\Gamma[\psi_3(2D)\to \chi_{c2}(2P)\gamma]$, and
$\Gamma[\eta_{c2}(2D)\to h_{c}(2P)\gamma]$ are relatively large (
about 100s keV). It is possible to establish the missing states
$\psi_2(2D)$, $\psi_3(2D)$ and $\eta_{c2}(2D)$ in the radiative
decay chains $\psi_2(2D)\to \chi_{c1}(2P)\gamma\to \psi(2S)\gamma
\gamma$, $\psi_3(2D)\to \chi_{c2}(2P)\gamma\to \psi(2S)\gamma
\gamma$, and $\eta_c(2^1D_2) \to h_{c}(2P)\gamma\to
\eta_c(1S)\gamma\gamma$, respectively.


The radiative decay properties of the $3P$ charmonium states are
very sensitive to their masses. The EM decays of $\chi_{cJ}(3P)$ and
$h_{c}(3P)$ are governed by the $\psi(3S)\gamma$ and
$\eta(3S)\gamma$ channels, respectively. The partial widths of the
$3P$ charmonium states decaying into the $2S$ states are also
sizeable. The partial decay widths for the transitions
$\chi_{cJ}(3P)\to \psi(3S)\gamma$ and $h_{c}(3P)\to \eta(3S)\gamma$
are about $10s-100s$ of keV. Thus, the observations of the missing
$3P$ charmonium states in the $\psi(3S)\gamma$ and $\eta(3S)\gamma$
channels are necessary in experiments. Furthermore, taking the newly
observed resonance $X(4350)$ as $\chi_{c2}(3P)$ as suggested by Liu
\emph{et al.}~\cite{Liu:2009fe}, we find that the branching ratios
of $\mathcal{B}[X(4350)\to\psi(3S)\gamma]$ and
$\mathcal{B}[X(4350)\to\psi(2S)\gamma]$ are $\sim 10^{-2}$ and $\sim
10^{-3}$, respectively. The $X(4350)$ resonance is most likely to be
observed in the $\psi(2S,3S)\gamma$ channels. We also find $X(4350)$
has a large branching ratio $\mathcal{B}[X(4350) \to
\psi_3(2D)\gamma]\simeq 2.4\times 10^{-3}$, thus, $X(4350)$ might
provide us a good source to look for the missing $\psi_3(2D)$ via
its radiative transition.

Finally, it should be pointed out that the harmonic oscillator wave
functions used by us might bring some uncertainties to our
predictions. Furthermore, the couple-channel effects (i.e., the
intermediate hadronic meson loops) might provide important
corrections to the EM decay properties of the charmonium states
lying above the open charm thresholds~\cite{Zhao:2013jza}, which
might affect our conclusions of the higher charmonium states. Our
quark model approach is also used to study the EM decay properties
of bottomonium states. For clarity, our results about the
bottomonium states will be reported in another work.

\section*{  Acknowledgement }

This work is supported, in part, by the National Natural Science
Foundation of China (Grants No. 11075051, No. 11375061, and No.
11405053), and the Hunan Provincial Natural Science Foundation
(Grant No. 13JJ1018).

\begin{table*}[htb]
\caption{Partial widths (keV) of the M1 radiative transitions for
some low-lying $S$-wave charmonium states. }\label{EM2}
\begin{tabular}{c|c|cccc|cccc|cc|cc}  \hline\hline
 Initial        & Final  & \multicolumn{4}{|c|} {\underline{$E_{\gamma}$ (MeV)}} & \multicolumn{4}{|c|} {\underline{$\Gamma_{\mathrm{M1}}$ (keV)}} & \underline{$\Gamma$ (keV)}  \\
   state             & state            & Ref.\cite{Ebert:2003}& NR\cite{Swanson:2005}&  GI \cite{Swanson:2005}& Ours & Ref.\cite{Ebert:2003}& NR& GI &  Ours &Exp. \\
\hline
$J/\psi$          &$\eta_{c}(1S)$        & 115 & 116 & 115 & 111    &1.05 & 2.9 & 2.4  &   1.25         &$1.58\pm0.37$\\
\hline
$\psi(2S)$        &$\eta_{c}(2S)$        & 32~ &48   & 48  & ~47    &0.043& 0.21& 0.17 & 0.10          &$0.21\pm0.15$\\
                  &$\eta_{c}(1S)$        & 639 & 639 & 638 & 635    &0.95 & 4.6 & 9.6  &   2.41          &$1.24\pm0.29$ \\
$\eta_{c}(2S)$    &$J/\psi$              & 514 & 501 & 501 & 502    &1.53 & 7.9 & 5.6  &   1.64          &  \\
\hline
$\psi(3S)$        &$\eta_{c}(3S)$        &     & 29  & 35~ & ~99    &   & 0.046 & 0.067 & 0.878         &  \\
                  &$\eta_{c}(2S)$        &     & 382 & 436 & 381    &   & 0.61  & 2.6   &  0.279         &  \\
                  &$\eta_{c}(1S)$        &     & 922 & 967 & 918    &   & 3.5   & 9.0   &  0.445         &  \\
\hline\hline
\end{tabular}
\end{table*}

\begin{table*}[htb]
\caption{Partial widths (keV) of the radiative transitions between the established charmonium states. For
comparison,  the predictions from the relativistic quark
model~\cite{Ebert:2003}, NR and GI models~\cite{Swanson:2005} and
SNR model~\cite{Li:2009zu} are listed in the table as well. The
experimental average data are taken from the PDG.
$\Gamma_{\mathrm{E1}}$ and $\Gamma_{\mathrm{EM}}$ stands for the E1
and EM transition widths, respectively. }\label{EMPW1}
\begin{tabular}{c|c|cccc|cccc|cc|cc}  \hline\hline
 Initial        & Final  & \multicolumn{4}{|c|} {\underline{$E_{\gamma}$ (MeV)}} & \multicolumn{4}{|c|} {\underline{$\Gamma_{\mathrm{E1}}$ (keV)}} & \multicolumn{2}{|c|} {\underline{$\Gamma_{\mathrm{EM}}$ (keV)}}  \\
   state             & state            & Ref.~\cite{Ebert:2003}& NR/GI~\cite{Swanson:2005}&  SNR~\cite{Li:2009zu}& Ours & Ref.\cite{Ebert:2003}& NR/GI~\cite{Swanson:2005}& SNR$_{0/1}$~\cite{Li:2009zu} & Ours & Ours &Exp. \\
\hline
$\psi(2S)$                      &$\chi_{c2}(1P)$        &128  & 128 /128 & 128 &128  & 18.2 & 38         / 24        &43/34 & 15    & 14 &$25.2\pm 2.9$  \\
                                &$\chi_{c1}(1P)$        &171  & 171 /171 & 171 &171  & 22.9 & 54         / 29        &62/36 & 20    & 22 &$25.5\pm 2.8$  \\
                                &$\chi_{c0}(1P)$        &259  & 261 /261 & 261 &261  & 26.3 & 63         / 26        &74/25 & 22    & 26 &$26.3\pm 2.6$  \\
\hline
$\chi_{c2}(1P)$           &      $J/\psi$               & 430 & 429 /429 & 429 & 429 & 327  & 424        / 313       &473/309 & 404   & 467  &$371\pm 34$\\
$\chi_{c1}(1P)$           &                             & 389 & 390 /389 & 390 & 390 & 265  & 314        / 239       &354/244 & 313   & 275  &$285\pm14$\\
$\chi_{c0}(1P)$           &                             & 305 & 303 /303 & 303 & 303 & 121  & 152        / 114       &167/117 & 159   & 128  &$133\pm8$\\
\hline
$h_{c}(1P)$                     &$\eta_{c}(1S)$         & 504 & 504 /496 & 504 & 499 & 560  & 498        / 352       &764/323 & 587   & 587  &$357\pm280$   \\
\hline
$\psi_1(1D)$                    &$\chi_{c2}(1P)$        & 234 & 208/208 & 213 & 215    & 6.9 & 4.9/3.3  & 5.8/4.6   & 3.4     & 2.6   &$<17.4$\\
                                &$\chi_{c1}(1P)$        & 277 & 250/251 & 255 & 258    & 135 & 125/77   & 150/93    & 83      & 70    &$81\pm 27$\\
                                &$\chi_{c0}(1P)$        & 361 & 338/338 & 343 & 346    & 355 &  403/213 & 486/197   & 245     &  218   &$202\pm 42$\\
$\psi_{2}(1D)$                  &$\chi_{c2}(1P)$        & 248 & 236/272 & 234 & 258    & 59  &  64/66   & 70/55     & 50      & 42     &  \\
                                &$\chi_{c1}(1P)$        & 291 & 278/314 & 276 & 299    & 215 & 307/268  & 342/208   & 226     &  227   &  \\
\hline
$\psi_1(2D)$                    &$\chi_{c2}(1P)$        &     & 559/590 &     & 587    &     &0.79/0.027&           & 0.46     & 8.8   &  \\
                                &$\chi_{c1}(1P)$        &     & 598/628 &     & 625    &     & 14 / 3.4 &           & 10     & 6.5   &  \\
                                &$\chi_{c0}(1P)$        &     & 677/707 &     & 704    &     & 27 / 35  &           & 27      & 21    &  \\
                                &$\chi_{c2}(2P)$        &     & 183/210 &     & 256    &     & 5.9/ 6.3 &           & 7.2     & 5.9   &    \\
\hline
$\eta_{c}(2S)$                 &$h_{c}(1P)$             &128  & 111 /119 & 109 &112  & 41   & 49 / 36   & 146/104   & 18    & 18 &              \\
\hline
$\psi(3S)$                      &$\chi_{c2}(2P)$        &     & 67  /119 & 119 &111  &      & 14         / 48        & &20     & 19 &              \\
                                &$\chi_{c2}(1P)$        &     & 455 /508 & 508 &455  &      & 0.70       / 13        & & 2.6   & 2.2 &              \\
                                &$\chi_{c1}(1P)$        &     & 494 /547 & 547 &494  &      & 0.53       / 0.85      & & 2.6   & 3.0 &              \\
                                &$\chi_{c0}(1P)$        &     & 577 /628 & 628 &577  &      & 0.27       / 0.63      & & 2.2   & 3.1 &              \\
\hline
$\chi_{c2}(2P)$                 &$\psi_{2}(1D)$         &     & 168 /139 & 139 & 103 &      & 17         / 5.6       & & 1.4   & 1.5&  \\
                                &$\psi_1(1D)$           &     & 197 /204 & 204 & 146 &      & 1.9        / 1.0       & & 0.26  & 0.32&  \\
                                &$\psi(2S)$             &     & 276 /282 & 235 & 234 &      & 304        / 207       &225/100 & 123   & 133  &  \\
                                &$J/\psi$               &     & 779 /784 & 744 & 742 &      & 81         / 53        &101/109 & 26    & 34 &  \\
\hline
\end{tabular}
\end{table*}

\begin{table*}[htb]
\caption{ Partial widths (keV) for the radiative transitions whose
initial and/or final charmonium states have not been established. In
the calculations, the masses of these well-established states are
taken the average values from the PDG. While, the masses of the
missing charmonium states are adopted the NR, GI and SNR potential
model predictions, which correspond to our three quark model
predictions QM$_1$, QM$_2$ and QM$_3$, respectively, for each
radiative transition. For comparison, the predictions from the NR
and GI models~\cite{Swanson:2005} and SNR model~\cite{Li:2009zu} are
listed in the table as well. }\label{EMPA}
\begin{tabular}{c|c|ccc|ccc|c|c}  \hline\hline
 Initial        & Final  & \multicolumn{3}{|c|} {\underline{$E_{\gamma}$ (MeV)}} & \multicolumn{3}{|c|} {\underline{$\Gamma_{\mathrm{E1}}$ (keV)}} & \multicolumn{1}{|c|} {\underline{$\Gamma_{\mathrm{EM}}$ (keV)}}  \\
   state             & state            &  NR/GI~\cite{Swanson:2005}&  SNR~\cite{Li:2009zu}& QM$_{1/2/3}$ &  NR/GI~\cite{Swanson:2005}& SNR$_{0/1}$~\cite{Li:2009zu} & ~~QM$_1$~~~~QM$_2$~~~~QM$_3$~~ & ~~QM$_1$~~~~QM$_2$~~~~QM$_3$~~  \\
\hline
$\psi(3S)$                      &$\chi_{c1}(2P)$             & 113 /145 &     &113/86/137       & 39    / 43     &        & 12~~~~~~~5.5~~~~~~21~~    & 13~~~~~~~5.7~~~~~~22~~      \\
                                &$\chi_{c0}(2P)$             & 184 /180 &     &184/122/193      & 54    / 22     &        & 16~~~~~~~5.1~~~~~~19~~    & 18~~~~~~~5.5~~~~~~21~~      \\
$\eta_{c}(3S)$                  &$h_{c}(2P)$                 & 108 /108 &     &108/107/82       & 105   / 64     &        & 32~~~~~~~31~~~~~~~14~~    & 32~~~~~~~31~~~~~~~14~~      \\
                                &$h_{c}(1P)$                 & 485 /511 &     &485/503/439       & 9.1   / 28    &        & 6.9~~~~~~8.7~~~~~~3.6~    & 6.9~~~~~~8.7~~~~~~3.6~      \\
\hline
$\chi_{c2}(2P)$                 &$\psi_{3}(1D)$              & 163 /128 &     & 119/77/126      & 88    / 29     &        & 12.0~~~~~3.4~~~~~~14.1    & 11.7~~~~~3.3~~~~~~13.8    \\
$\chi_{c1}(2P)$                 &$\psi_{2}(1D)$              & 123 /113 &     & 101/128/77      & 35    / 18     &        & 6.6~~~~~~13~~~~~~3.0~    & 6.6~~~~~~13~~~~~~3.0~    \\
                                &$\psi_1(1D)$                & 152 /179 &     & 144/171/121     & 22    / 21     &        & 6.2~~~~~~10~~~~~~~3.7~    & 6.8~~~~~~11~~~~~~~4.1~    \\
                                &$\psi(2S)$                  & 232 /258 & 182 & 232/258/209     & 183   / 183    & 103/60 & 120~~~~~~161~~~~~~90~~    & 111~~~~~~148~~~~~~84~~     \\
                                &$J/\psi$~~~~~               & 741 /763 & 697 & 741/763/721     & 71    / 14     & 83/45  & 26~~~~~~~31~~~~~~~22~~    & 20~~~~~~~24~~~~~~~17~~     \\
$\chi_{c0}(2P)$                 &$\psi_1(1D)$                & 81  /143 &     & 73/136/63       & 13    / 51     &        & 3.4~~~~~~21~~~~~~~2.2~    & 3.5~~~~~~22~~~~~~~2.3~    \\
                                &$\psi(2S)$                  & 162 /223 & 152 & 257/223/248     & 64    / 135    & 61/44  & 159~~~~~~108~~~~~~144~    & 133~~~~~~93~~~~~~~121~    \\
                                &$J/\psi$                    & 681 /733 & 672 & 681/733/673     & 56    / 1.3    & 74/9.3 & 16~~~~~~~24~~~~~~~15~~    & 9.4~~~~~~14~~~~~~~8.8~    \\
$h_{c}(2P)$                     &$\eta_{c2}(1D)$             & 133 /117 &     & 133/117/110     & 60    / 27     &        & 20~~~~~~~14~~~~~~~11~~    & 20~~~~~~~14~~~~~~~11~~     \\
                                &$\eta_{c}(2S)$              & 285 /305 & 261 & 284/304/260     & 280   / 218    & 309/108& 208~~~~~~249~~~~~~164~    & 208~~~~~~249~~~~~~164~    \\
                                &$\eta_{c}(1S)$              & 839 /856 & 818 & 835/853/815     & 140   / 85     & 134/250& 51~~~~~~~58~~~~~~~45~~    & 51~~~~~~~58~~~~~~~45~~    \\
\hline
$\chi_{c2}(3P)$                 &$\psi_{3}(2D)$              & 147 /118 &     &147/118/104        & 148   /  51    &        & 44~~~~~~~23~~~~~~~16~~~    & 43~~~~~~~23~~~~~~~16~~~     \\
                                &$\psi_{2}(2D)$              & 156 /127 &     &156/127/107        & 31    /  10    &        & 9.2~~~~~~5.1~~~~~~3.1~~~    & 10~~~~~~5.6~~~~~~3.3~~~     \\
                                &$\psi_1(2D)$                & 155 /141 &     &124/144/17       & 2.1   /  0.77  &        & 0.32~~~~~0.48~~~~~0~~~    & 0.38~~~~~0.60~~~~~0~~~     \\
                                &$\psi_{3}(1D)$              & 481 /461 &     &481/461/389      & 0.049 /  6.8   &        & 4.8~~~~~~3.6~~~~~~1.3~    & 5.5~~~~~~4.2~~~~~~1.5~   \\
                                &$\psi_{2}(1D)$              & 486 /470 &     &466/484/367      & 0.01  /  0.13  &        & 0.70~~~~~~0.88~~~~~~0.15    & 2.1~~~~~~2.7~~~~~~0.47  \\
                                &$\psi_1(1D)$                & 512 /530 &     &505/523/408      & 0.00  /  0.00  &        & 0.08~~~~~~0.09~~~~~~0.02    & 1.9~~~~~~2.3~~~~~~0.54  \\
                                &$\psi(3S)$                  & 268 /231 &     &268/287/165      & 509   /  199   &        & 235~~~~~~278~~~~~~63~~    & 257~~~~~~306~~~~~~66~~    \\
                                &$\psi(2S)$                  & 585 /602 &     &585/602/490      & 55    /  30    &        & 24~~~~~~~29~~~~~~~8.2~    & 30~~~~~~~35~~~~~~~9.7~   \\
                                &$J/\psi$                    & 1048/1063&     &1048/1063/964    & 34    /  19    &        & 4.6~~~~~~5.2~~~~~~2.1~    & 6.5~~~~~~7.5~~~~~~3.0~   \\
$\chi_{c1}(3P)$                 &$\psi_{2}(2D)$              & 112 /108 &     & 112/108/77       & 58    /  35    &        & 18~~~~~~~16~~~~~~~6.0~~~    & 18~~~~~~~16~~~~~~~6.0~~~     \\
                                &$\psi_1(2D)$                & 111 /121 &     & 79/124/0        & 19    /  15    &        & 2.2~~~~~~8.0~~~~~~0~~~    & 2.3~~~~~~8.7~~~~~~0~~~     \\
                                &$\psi_{2}(1D)$              & 445 /452 &     & 425/466/340     & 0.035 /  4.6   &        & 2.0~~~~~~3.5~~~~~~0.47    & 2.5~~~~~~4.4~~~~~~0.60  \\
                                &$\psi_1(1D)$                & 472 /512 &     & 465/505/381     & 0.014 /  0.39  &        & 1.1~~~~~~1.9~~~~~~0.32    & 1.5~~~~~~2.6~~~~~~0.41  \\
                                &$\psi(3S)$                  & 225 /212 &     & 225/268/136     & 303   /  181   &        & 148~~~~~~235~~~~~~36~~    & 137~~~~~~215~~~~~~35~~    \\
                                &$\psi(2S)$                  & 545 /585 &     & 545/585/463     & 45    /  8.9   &        & 16~~~~~~~24~~~~~~~5.8~    & 13~~~~~~~20~~~~~~~5.0~   \\
                                &$J/\psi$                    & 1013/1048&     &1013/1048/941    & 31    /  2.2   &        & 3.4~~~~~~4.6~~~~~~1.7~    & 2.4~~~~~~3.2~~~~~~1.3~   \\
$\chi_{c0}(3P)$                 &$\psi_1(2D)$                & 43  /97  &     &  11/100/0       & 4.4   /  35    &        & 0.02~~~~~17~~~~~~~0~~~    & 0.02~~~~~18~~~~~~~0~~~     \\
                                &$\psi_1(1D)$                & 410 /490 &     & 403/483/338     & 0.037 /  9.7   &        & 1.9~~~~~~5.8~~~~~~0.60    & 2.1~~~~~~6.8~~~~~~0.66  \\
                                &$\psi(3S)$                  &159  /188 &     & 159/245/90      & 109   /  145   &        & 57~~~~~~~185~~~~~~11~~    & 51~~~~~~~156~~~~~~10~~    \\
                                &$\psi(2S)$                  &484  /563 &     & 484/563/421     & 32    /  0.045 &        & 7.7~~~~~~19~~~~~~~3.2~    & 5.4~~~~~~13~~~~~~~2.3~   \\
                                &$J/\psi$                    &960  /1029&     &960/1029/905     & 27    /  1.5   &        & 2.1~~~~~~3.9~~~~~~1.2~    & 0.95~~~~~1.7~~~~~~0.58  \\
$h_{c}(3P)$                     &$\eta_{c2}(2D)$             &119  /109 &     & 119/109/84       & 99    /  48    &        & 28~~~~~~~22~~~~~~~10~~~    & 28~~~~~~~22~~~~~~~10~~~     \\
                                &$\eta_{c2}(1D)$             &453  /454 &     & 453/454/370     & 0.16  /  5.7   &        & 3.9~~~~~~4.0~~~~~~1.1~    & 6.3~~~~~~6.4~~~~~~1.8~   \\
                                &$\eta_{c}(3S)$              &229  /246 &     & 229/247/189     & 276   /  208   &        & 156~~~~~~189~~~~~~92~~    & 156~~~~~~189~~~~~~92~~    \\
                                &$\eta_{c}(2S)$              &593  /627 &     & 592/626/510     & 75    /  43    &        & 39~~~~~~~54~~~~~~~16~~    & 39~~~~~~~54~~~~~~~16~~    \\
                                &$\eta_{c}(1S)$              &1103 /1131&     &1099/1128/1028   & 72    /  38    &        & 6.9~~~~~~8.6~~~~~~3.8~    & 6.9~~~~~~8.6~~~~~~3.8~   \\
\hline\hline
\end{tabular}
\end{table*}


\begin{table*}[htb]
\caption{Partial widths (keV) for the radiative transitions of
unestablished $D$-wave charmonium states. In the calculations, the
masses of these well-established states are taken the average values
from the PDG. While, the masses of the missing charmonium states are
adopted the NR, GI and SNR potential model predictions, which
correspond to our three quark model predictions QM$_1$, QM$_2$ and
QM$_3$, respectively, for each radiative transition. For comparison,
the predictions from the relativistic quark model~\cite{Ebert:2003},
NR and GI models~\cite{Swanson:2005}, and SNR model~\cite{Li:2009zu}
are listed in the table as well. }\label{D-wave}
\begin{tabular}{c|c|cccc|cccc|cc|cc}  \hline\hline
 Initial      & Final  & \multicolumn{4}{|c|} {\underline{$E_{\gamma}$ (MeV)}} & \multicolumn{4}{|c|} {\underline{$\Gamma_{\mathrm{E1}}$ (keV)}} & \multicolumn{1}{|c} {\underline{$\Gamma_{\mathrm{EM}}$ (keV)}}  \\
   state             & state            & \cite{Ebert:2003}& NR/GI~\cite{Swanson:2005}&  SNR~\cite{Li:2009zu} & QM$_{1/2/3}$ & \cite{Ebert:2003}& NR/GI~\cite{Swanson:2005} & SNR$_{0/1}$~\cite{Li:2009zu} & ~~QM$_1$~~~~QM$_2$~~~~QM$_3$~~  & ~~QM$_1$~~~~QM$_2$~~~~QM$_3$~~ \\

\hline
$\psi_{3}(1D)$                  &$\chi_{c2}(1P)$        & 250 & 242/282 & 236 & 242/282/235    & 156 & 272 /296  & 284/223   & 167~~~~~256~~~~~155~    & 175~~~~~271~~~~~162    \\
                                &$\chi_{c1}(1P)$        &     &         &     & 284/323/277    &     &           &           & 0~~~~~~~0~~~~~~~0    & 0.54~~~~1.1~~~~0.48   \\
                                &$\chi_{c0}(1P)$        &     &         &     & 371/410/365    &     &           &           & 0~~~~~~~0~~~~~~~0    & 0.60~~~~1.2~~~~0.53   \\
$\eta_{c2}(1D)$                 &$h_{c}(1P)$            & 275 & 264/307 & 260 & 264/299/261    & 245 & 339 /344  & 575/375   & 214~~~~~302~~~~~208~    & 214~~~~~302~~~~~208   \\
\hline
$\psi_{3}(2D)$                  &$\chi_{c2}(1P)$        &     & 566/609 &     & 566/609/511    &     &  29 / 16  &           & 13~~~~~~21~~~~~~7.0~~    & 16~~~~~~25~~~~~~8.5    \\
                                &$\chi_{c1}(1P)$        &     &         &     & 604/647/549    &     &           &           & 0~~~~~~~0~~~~~~~0~~    & 6.2~~~~~9.1~~~~~3.6    \\
                                &$\chi_{c0}(1P)$        &     &         &     & 684/726/630    &     &           &           & 0~~~~~~~0~~~~~~~0~~    & 6.2~~~~~8.5~~~~~4.0    \\
$\psi_{2}(2D)$                  &$\chi_{c2}(1P)$        &     & 558/602 &     & 558/601/508    &     & 7.1 /0.62 &           & 3.1~~~~~4.8~~~~~1.7~~    & 6.7~~~~~10~~~~~~3.8    \\
                                &$\chi_{c1}(1P)$        &     & 597/640 &     & 597/639/547    &     &  26 / 23  &           & 14~~~~~~21~~~~~~8.0~~    & 17~~~~~~26~~~~~~10    \\
$\eta_{c2}(2D)$                 &$h_{c}(1P)$            &     & 585/634 &     & 585/628/534    &     &  40 / 25  &           & 16~~~~~~25~~~~~~9.2~~    & 26~~~~~~39~~~~~~15    \\
\hline
$\psi_{3}(2D)$                  &$\chi_{c2}(2P)$        &     & 190/231 &     & 233/280/172    &     & 239 /272  &           & 202~~~~~330~~~~~87~    & 212~~~~~349~~~~~90   \\
                                &$\chi_{c1}(2P)$        &     &         &     & 235/256/197    &     &           &           & 0.24~~~~0.42~~~~0.07~    & 0.44~~~~0.72~~~~0.16    \\
                                &$\chi_{c0}(2P)$        &     &         &     & 303/290/253    &     &           &           & 0.65~~~~0.49~~~~0.19~~    & 0.70~~~~0.52~~~~0.20    \\
$\psi_{2}(2D)$                  &$\chi_{c2}(2P)$        &     & 182/223 &     & 225/272/169    &     & 52  / 65  &           & 46~~~~~~76~~~~~~21~    & 39~~~~~~64~~~~~~19   \\
                                &$\chi_{c1}(2P)$        &     & 226/247 &     & 226/247/194    &     & 298 /225  &           & 140~~~~~178~~~~~92~    & 140~~~~~178~~~~~92   \\
$\psi_1(2D)$                    &$\chi_{c1}(2P)$        &     & 227/234 &     & 258/231/280    &     & 168 /114  &           & 110~~~~~~82~~~~~~137~    & 92~~~~~~70~~~~~~113   \\
                                &$\chi_{c0}(2P)$        &     & 296/269 &     & 325/266/334    &     & 483 /191  &           & 268~~~~~162~~~~~287~    & 240~~~~~146~~~~~256   \\
$\eta_{c2}(2D)$                 &$h_{c}(2 P)$           &     & 218/244 &     & 218/244/187    &     & 336 /296  &           & 168~~~~~230~~~~~109~    & 169~~~~~231~~~~~109   \\
\hline\hline
\end{tabular}
\end{table*}

\end{document}